\documentclass[lettersize,journal]{IEEEtran}
\renewcommand{\t}{^{\mbox{\tiny {T}}}}

\newcommand{\upfrown}[1]{
	\mathord{\buildrel{\lower3pt\hbox{$\scriptscriptstyle\frown$}} 
		\over #1}
}

\newcommand{\eproof}{\hfill\rule{2mm}{2mm}}

\newcommand{\bstate}{\medskip\begin{state} }
	\newcommand{\estate}{ \hfill  \rule{1mm}{2mm}\medskip\end{state}}

\newcommand{\bass}{\medskip\begin{ass} }
	\newcommand{\eass}{ \hfill  \rule{1mm}{2mm}\medskip\end{ass}}

\newcommand{\brem}{\medskip \begin{remark}  }
	\newcommand{\erem}{\hfill \rule{1mm}{2mm}\medskip
\end{remark} }
\newcommand{\bthm}{\medskip\begin{theorem}  }
	\newcommand{\ethm}{ \hfill  \rule{1mm}{2mm} \medskip
\end{theorem} }
\newcommand{\blem}{\medskip\begin{lemma}  }
	\newcommand{\elem}{ \hfill \rule{1mm}{2mm}\medskip
\end{lemma} }
\newcommand{\bcorollary}{\medskip\begin{corollary}  }
	\newcommand{\ecorollary}{  \hfill \rule{1mm}{2mm}\medskip
\end{corollary} }
\newcommand{\bdefn}{\medskip\begin{definition}}
	\newcommand{\edefn}{  \hfill \rule{1mm}{2mm}\medskip
\end{definition} }
\newcommand{\bproposition}{\medskip\begin{proposition} }
	\newcommand{\eproposition}{\hfill \rule{1mm}{2mm}\medskip
\end{proposition} }
\newcommand{\bexample}{\medskip\begin{example} \rm}
	\newcommand{\eexample}{ \hfill \rule{1mm}{2mm}\medskip
\end{example} }

\newcommand{\bcon}{\medskip\begin{condition} \rm}
	\newcommand{\econ}{ \hfill \rule{1mm}{2mm}\medskip
\end{condition} }

\newcommand{\prooflater}[1]{\noindent{\bf Proof of #1: }}

\newtheorem{theorem}{\bf Theorem}[section]
\newtheorem{ass}{\bf Assumption}[section]
\newtheorem{lemma}{\bf Lemma}[section]
\newtheorem{definition}{\bf Definition}[section]
\newtheorem{remark}{\bf Remark}[section]
\newtheorem{corollary}{\bf Corollary}[section]
\newtheorem{proposition}{\bf Proposition}[section]
\newtheorem{example}{\bf Example}[section]
\newtheorem{condition}{\bf Condition}[section]
\newtheorem{state}{\bf Assumption}[section]

\usepackage{color}
\usepackage{amsmath,amsfonts}
\usepackage{subfigure}

\usepackage{verbatim}
\usepackage{graphicx}
\usepackage{ulem}

\begin{document}

\title{An Improved Active Disturbance Rejection Control  for Bode's Ideal Transfer Function}
\author{Bolin Li, and  Lijun Zhu}
\maketitle

\begin{abstract}
This paper presents an  active disturbance rejection control (ADRC) scheme with an improved fractional-order extended state observer (IFO-ESO). 
Based on the new ADRC scheme, 
the   open-loop transfer function of a high-order system can be approximately rendered to  
a so-called Weighed Bode's ideal transfer function, whose closed-loop performance is less prone to the controller parameter variations.
 The design of the IFO-ESO helps reduce the number of system states to be estimated and improves the performance of  closed-loop system over 
the fractional-order active disturbance rejection control (FO-ADRC) in the literature. Compared with the integer-order active disturbance rejection controller (IO-ADRC) and FO-ADRC, the auxiliary tracking controller of IFO-ADRC has a simpler form. Frequency-domain analysis shows that IFO-ESO has better estimation performance than typical fractional-order ESO (FO-ESO), and time-domain simulation demonstrates that the proposed ADRC has better transient performance and is more robust against the parameter variations than FO-ADRC and IO-ADRC. 
The proposed ADRC is applied to permanent magnet synchronous motor (PMSM) servo control system and demonstrates its capability in a real-world application.
\end{abstract}


\section{Introduction}
Fractional calculus has been applied for the controller design
in the fields of  permanent magnet synchronous motor  gas-turbine  and  heating–furnace etc.,
to improve the system performance 
 in recent years \cite{liu2018guaranteed,pu2016fractional,tzounas2020theory}.
 Many fractional-order controllers have been proposed, including fractional-order sliding mode controller \cite{zaihidee2019application},  intelligent PID controller \cite{gomaa2019novel},  PID controller \cite{li2009fractional},  active disturbance rejection controller \cite{li2016fractional}, and so on. 

In 1940s, Bode \cite{bode1945network}  suggested an ideal open-loop transfer function, called  Bode's ideal transfer function (BITF).
An advantage of  the BITF is that when the  open-loop gain varies,  the gain crossover frequency will change while
the phase margin remains the same.  
Due to this characteristic, 
 BITF has been used as the reference model to
  design the fractional-order controller  \cite{zheng2021synthesis,zhuo2020fractional,barbosa2004tuning}. Using the BITF method, the closed-loop system achieve the iso-damping property (the phase derivative with respect to the frequency is zero) and the overshoots of the  step responses will remain almost constant even when the open-loop  gain varies. The controller design   is straightforward  for  low-order systems by using BITF method \cite{jadhav2014robust,patil2015design,zhang2014fractional}, but  becomes complicated for high-order systems \cite{al2020robustness}, in particular with uncertainties.



The integer-order active disturbance rejection control (IO-ADRC) proposed by Gao \cite{gao2006scaling} provides a strategy for control system
design with uncertainties. The core idea of IO-ADRC is to improve the robustness of the system using extended state observer (IO-ESO, or extended high-gain observer) \cite{ran2021new, feng2014active}  to estimate the total disturbances, including internal disturbance caused by system uncertainties and external disturbance \cite{serrano2020composite,chen2016output}.   
Inspired by the ADRC structure, 
a fractional-order ADRC  (FO-ADRC) was proposed in \cite{al2020robustness}, and approximately   converted a class of integer-order systems into a weighted Bode's ideal transfer function (WBITF). WBITF is  a higher-order transfer function  and can be approximated to a BITF in the low-frequency band.


This paper is along the research line of proposing a new type of fractional-order ESO structure and then an improved fractional-order ADRC (IFO-ADRC) to achieve a WBITF. The main contributions of the paper can be summarized as follows. First, a new type of ESO, called IFO-ESO, is proposed to compensate for the system uncertainties and external disturbance. 
In comparison to \cite{al2020robustness},  less  state is estimated in the  IFO-ESO and high-order dynamics are not included in the extended state, which potentially improves the disturbance estimation performance of IFO-ESO over FO-ESO,  making the system more robust to  system uncertainties and external disturbances. 
 Better estimation performance of IFO-ESO ensures that the open-loop transfer function can be more accurately  approximated  into a newly introduced WBITF.
With proposed IFO-ESO, the auxiliary tracking controller of IFO-ADRC also has a simpler form. Second, the stability criteria of the IFO-ESO and the IFO-ADRC closed-loop system are given.
For a second-order system, we prove that when the observer gains are selected to  be sufficiently large, the closed-loop system is BIBO stable.

The rest of the paper is organized as follows. Some preliminaries on fractional operator and BITF are provided and the problem is formulated in Section \uppercase\expandafter{\romannumeral2}. The structures of  IFO-ADRC, and the BIBO stability criteria for the IFO-ESO and the closed-loop system are given in Section \uppercase\expandafter{\romannumeral3}. The performance analysis of the IFO-ESO in the frequency-domain is shown in Section \uppercase\expandafter{\romannumeral4}.
 Section \uppercase\expandafter{\romannumeral5} presents the time-domain simulation results, followed by the experimental results on PMSM servo system  in Section \uppercase\expandafter{\romannumeral6}. The paper is concluded in Section \uppercase\expandafter{\romannumeral7}.

\section{Preliminaries and Problem Formulation}
\subsection{Bode's Ideal Transfer Function}
The mathematical form  of   Bode's ideal transfer function \cite{bode1945network}  
 is  
\begin{gather}
L_o(s) = {\left( {\frac{{{\omega _g}}}{s}} \right)^\chi }
\label{Bode's ideal transfer function}
\end{gather}
where $\omega_g$ is the gain crossover frequency and $\chi$ is a real-number order. 
In Bode plot, the slope of the magnitude curve of BITF is -20$\chi$ dB/deg, and the phase curve is a horizontal line at $\chi\pi/2$ rad, both parameterized by $\chi$. An advantage of  the BITF is that when the  open-loop gain varies,  the crossover frequency $\omega_g$ will change while
the phase margin constant $ \pi (1-\chi/2)$ rad remains the same.  When $1<\chi<2$, the step response of the unit negative feedback system of a BITF  is similar to that of an under-damped second-order system.
Due to this characteristic, the BITF has been adopted as the reference model  for the controller design, particularly for
low-order systems. 
However,  converting a higher-order system  into  a BITF in (\ref{Bode's ideal transfer function}) requires an irrational compensator, and is not feasible in practice. 
 In \cite{al2020robustness}, a so-called weighted BITF (WBITF) was then proposed as the series combination of the BITF and $k$ low-pass filters,    
for $1<\chi<2$, 
\begin{gather}
{L_{ol}}(s) = \frac{{\omega _g^\chi }}{{{s^\chi }{{(Ts + 1)}^\kappa }}}
\label{fo-wbitf}
\end{gather}   
where $\kappa$ is a positive integer satisfying $\kappa + \chi = m$ and $m$ represents the maximum order of the plant, and $T$ is the time constant of the filter. Note that when $T$ is much less than $\omega_g$, the characteristic of (\ref{fo-wbitf}) is similar to that of  BITF  in (\ref{Bode's ideal transfer function}).  An  FO-ADRC was then constructed  to convert an high-order system into a WBITF with the same order.

\subsection{Problem Formulation}
 In this paper, we consider a high-order linear system as follows
\begin{equation}
G(s) = \frac{{Y(s)}}{{U(s)}} = \frac{b}{{{s^{m }} + \sum\limits_{i = 1}^{m - 1} {{a_i}{s^{i }}}  + {a_0}}}
\label{eq_6}
\end{equation}
where $s$ is Laplace operator, $a_i$, $a_0$ and $b$ are real numbers, $m$ and $i$ are positive integers with $m$ representing the maximum order of the system. 
The differential equation form of system (\ref{eq_6}) with the external disturbance, denoted by $d$, is
\begin{equation}
{y^{(m )}} =  - \mathop \sum \limits_{i = 1}^{m - 1} {a_i}{y^{(i)}} - {a_0}y + bu + d
\label{eq_7}
\end{equation}

The aim of this paper is first to  propose an improved FO-ESO based on which the system is   approximately converted  (\ref{eq_6}) into a  newly introduced WBITF   as follows 
\begin{gather}
{G_{ol}}(s) = \frac{{\omega _g^\chi }}{{{s^\chi }{{(Ts^{\gamma} + 1)}^\kappa }}}
\label{ifo-wbitf}
\end{gather}  
where $\gamma$ is the order of the fractional-order filter, $0<\gamma<2$, and $\chi+\kappa\gamma = m$. WBITF (\ref{ifo-wbitf}) is composed of a BITF and $\kappa$  low-pass filters in series with the order $\gamma$ to be specified. When $\gamma=1$, WBITF (\ref{ifo-wbitf}) coincides with WBITF (\ref{fo-wbitf}). In next section, the new WBITF with fractional-order $\gamma$ facilitates the IFO-ESO to be proposed in next section.

The second aim  of this paper is to design $u$ such that  the system output tracks a sufficiently smooth reference trajectory $r$ and the ultimate tracking error stays in the neighborhood of the origin,  i.e., $\lim_{t\rightarrow \infty} \|y(t)-r(t)\|<\epsilon$, when   the reference signal and its derivatives, i.e., $r$, $ \dot r,\ddot r, \cdots {r^{(m-1)}},{r^{(m)}}$ are  bounded. 



\subsection{Definitions on Fractional Operator}
There are various definitions of fractional derivative \cite{teodoro2019review}, such as the Grunwald–Letnikov, Riemann–Liouville, Caputo definitions, and so on. The Grunwald–Letnikov (GL) is one of the most commonly used definition and adopted in this paper, whose notation is given as \cite{yumuk2019analytical} 
\begin{gather}
{}_aD_t^\gamma f(t) = \mathop {\lim }\limits_{h \to 0} \frac{1}{{{h^\gamma }}}\sum\limits_{j = 0}^{[\frac{{t - a}}{h}]} {{{( - 1)}^j}\left( \begin{array}{l}
	n_0\\
	j
	\end{array} \right)} f(t - jh)
\end{gather}
where $n_0$ is an integer satisfying $n_0 - 1 < \gamma  < n_0$ and $\gamma$ is a fractional order, $h$ and $[\frac{t-a}{h}]$ are time increment and integer part of the upper limit of summation, respectively. The binomial coefficient   is  
$
\left( \begin{array}{l}
{n_0}\\
j
\end{array} \right) = \frac{{\Gamma ({n_0} + 1)}}{{\Gamma (j + 1)\Gamma ({n_0} - j + 1)}}
$
where $\Gamma ( \bullet )$ is Euler's gammafunction.  The Laplace transform of the GL fractional-order derivative with  zero initial condition is given as
$
\mathcal{L}{\{ _a}D_t^\gamma f(t)\}  = {s^\gamma }F(s).
$

\section{Improved Active disturbance rejection controls \label{sec:ADRC}}
\subsection{Structures of IFO-ESO and IFO-ADRC}

In this section, we will  use    IFO-ESO and IFO-ADRC to approximately convert the system (\ref{eq_6}) into a WBITF (\ref{ifo-wbitf}).
Equation (\ref{eq_7}) can be rewritten as follows,
\begin{equation}
{y^{(m)}} 
=   {f_{ifo}}({y^{(1)} },{y^{(2) }}, \cdots ,{y^{(m - 1) }},y,u,t) + {b_0}u{\mkern 1mu}
\label{eq_8}
\end{equation}
where $f_{ifo}=- \sum\limits_{i = 1}^{m - 1} {{a_i}} {y^{(i )}} - {a_0}y + (b - {b_0})u+d$. Note that  $f_{ifo}$ can be regarded as the total disturbance where the term $- \sum\limits_{i = 1}^{m - 1} {{a_i}} {y^{(i )}} - {a_0}y + (b - {b_0})u$ is the internal disturbance due to  uncertain parameters and $d$ is the external disturbance. 

Let $\chi$ be a fractional number satisfying $1<\chi<2$, $n = [\frac{m}{\chi}]+1$, $\gamma = \frac{m-\chi}{n-1}$, and $\gamma < \nu < \chi$. Note ${x_1} = y,{x_2} = {y^{(\chi)}},{x_3} = {y^{(\gamma + \chi)}}, \cdots ,{x_{n}} = {y^{((n-2)\gamma + \chi)}}, {x_{n+1}} = {f_{ifo}}, {h_{ifo}} = {{f}^{(\nu)}_{ifo}}({y^{(1)} },{y^{(2) }}, \cdots ,{y^{(m - 1) }},y,u,t)$ where $x_1, x_2, \cdots, {x_{n}}$ represent system states and ${x_{n+1}}$ is an extended state. Let $x = {\left[{\begin{array}{*{20}{c}}
			{{x_1}},{{x_2}}, \cdots ,{{x_n}},{{x_{n + 1}}}
	\end{array}} \right]\t}$. The state-space representation of (\ref{eq_8}) is given as follows
\begin{gather}
{}_aD_t^\textbf{q}x = Ax + Bu + E{h_{ifo}} \nonumber \\
{y = Cx}
\label{eq_9}
\end{gather}
where $q = {\left[{\begin{array}{*{20}{c}}
		\chi, \gamma, \cdots ,{\gamma},\nu
		\end{array}} \right]\t}$
	and 
\begin{gather}
A =\left[\begin{array}{cc}
0_{n\times1} & I_{n} \nonumber\\
0 & 0_{1\times n}
\end{array}\right],
B = {\left[{\begin{array}{*{20}{c}}
		0,0, \cdots ,{{b_0}},0
		\end{array}} \right]\t},\nonumber\\
C = \left[ {\begin{array}{*{20}{c}}
	1,0, \cdots ,0,0
	\end{array}} \right],
E = {[\begin{array}{*{20}{c}}
	0& \cdots &0&1
	\end{array}]\t}.
\end{gather}
Then,  IFO-ESO is designed to estimate  $x$ as follows
\begin{gather}
{}_aD_t^\textbf{q}z = Az + Bu + L(y - \upfrown{y}) \nonumber\\
\upfrown{y} = Cz
\label{eq_10}
\end{gather}
where $z = \left[{\begin{array}{*{20}{c}}
	{{z_1}},{{z_2}}, \cdots ,{{z_n}},{{z_{n + 1}}}
	\end{array}} \right]\t$ and
\begin{gather} 
L = \left[ {\begin{array}{*{20}{c}}
	{{\beta _1}},{{\beta _2}}, \cdots ,{{\beta _n}},{{\beta _{n + 1}}}
	\end{array}} \right]\t
\label{eq:ABC}
\end{gather}
Note that $L$ are extended state observer gains, $z$ is the estimation of the state$ x$ and $b_0$ is the nominal value of $b$.

\begin{figure}[!ht]
	\centering
	\includegraphics[scale=0.55]{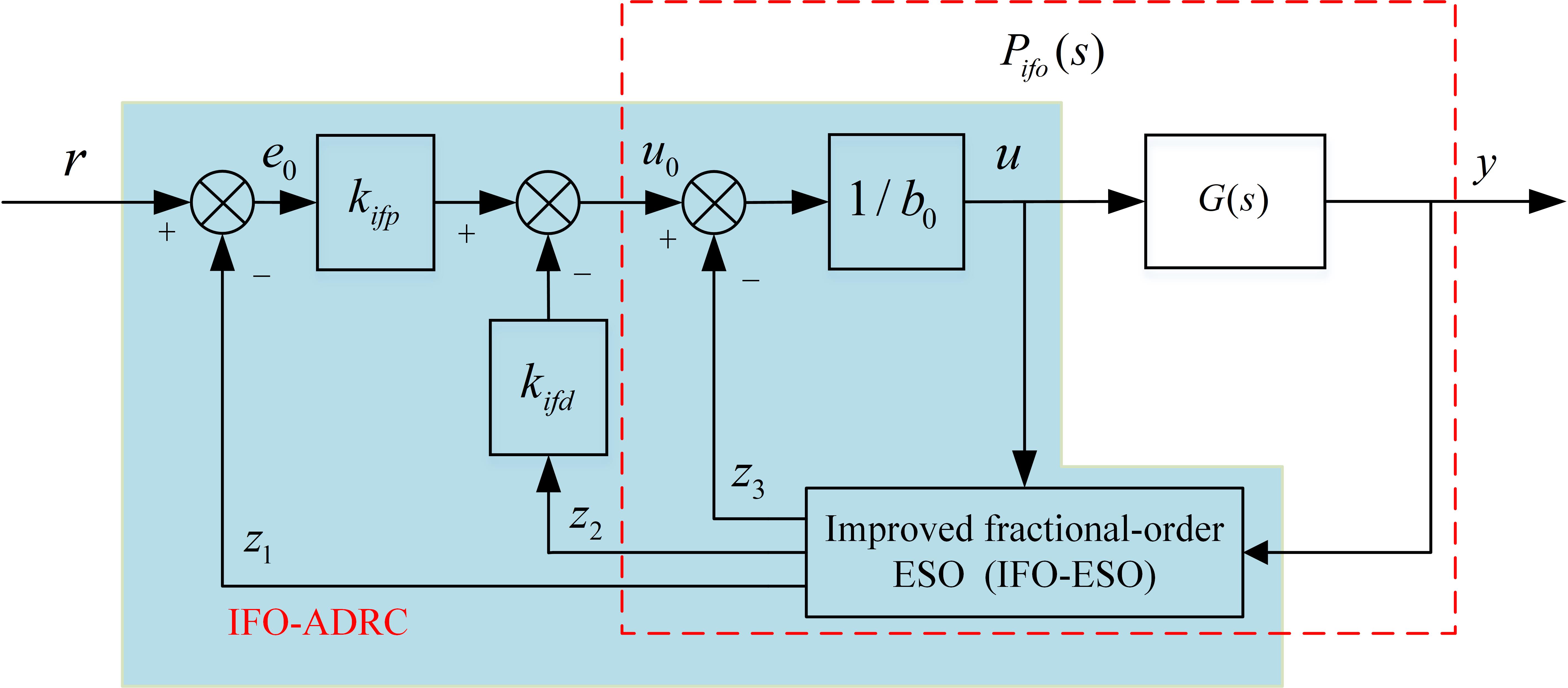}
	\caption{Structure of the IFO-ADRC for a second-order plant.}
	\label{str_IFO}
\end{figure}

The  controller is designed as follows
\begin{equation}
u = \frac{u_0  -\upfrown{f}_{ifo}}{b_0}
\label{eq_8_}
\end{equation}
where $\upfrown{f}_{ifo}	= {z_{n+1}}$ and $u_0$ is auxiliary tracking controller to be designed. Then, the closed-loop system  composed of (\ref{eq_8}) and (\ref{eq_8_})  becomes 
\begin{equation}
{y^{(m)}} = {u_0}  + (f_{ifo} - \upfrown{ f}_{ifo}).
\label{eq_14}
\end{equation}
The tracking task can be fulfilled with 
the auxiliary tracking controller $u_0 $ in (\ref{eq_8_}), designed as follows,
\begin{gather}
{u_0} = {k_p}e_0 - K_0z + K_1 \hat r 
\label{eq_12}
\end{gather}
where $e_0$ = $r-z_1$, $\hat r = [{r_1},{r_2}, \cdots ,{r_n},{r_{n + 1}}]$ with $r_1 = r$, $r_i = r^{(\chi+(i-2)\gamma)}$ for $i =2,3,\cdots,n+1$. Note that $r$ is the reference input of the closed-loop system. 
In particular, we choose the controller gain as $
K_0 = \left[{\begin{array}{*{20}{c}}
	{{0}},{{k_{d_{1}}}}, \cdots ,{{k_{d_{n-1}}},{0}}
	\end{array}} \right]$ with  ${k_{{d_i}}} = C_{n-1}^i{\omega _c}^{n -1- i}
$ for $i =1,2,\cdots,n-1$ and $
K_1 = \left[{\begin{array}{*{20}{c}}
	{{0}},{{k_{d_{1}}}}, \cdots ,{{k_{d_{n-1}}},{1}}
	\end{array}} \right]$. 
Substituting (\ref{eq_12}) into (\ref{eq_14})   gives
\begin{equation}
{y^{((n - 1)\gamma  + \chi )}} + K_0x 
 = {k_p}{e_0} + K_0(x - z) + ({f_{ifo}} - {\upfrown{f}}_{ifo}) + K_1 \hat r 
\label{open_loop-1}
\end{equation}
where we note  $m=(n-1)\gamma + \chi$. 
The structure of the IFO-ADRC for a second-order plant with  $\dot{r} = 0$ and $\ddot{r} = 0$,  is illustrated in Fig.~\ref{str_IFO}.

Now, let us consider the open-loop transfer function of the IFO-ADRC and without loss of generality  assume $r\equiv 0$.
If signals $f_{ifo}$ and $x$ are well estimated by  $\upfrown{f}_{ifo}$ and $z$, respectively, i.e., $\upfrown{f}_{ifo}\approx f_{ifo}$ and $z \approx x$, the Laplace transform of the both side of  (\ref{open_loop-1}) (with $r\equiv 0$) gives
\begin{gather}
\frac{{Y(s)}}{{{E_0}(s)}} \approx k_p\frac{{\frac{1}{{{\omega _c}^{n - 1}}}}}{{{s^\chi }{{(\frac{1}{{{\omega _c}}}{s^\gamma } + 1)}^{n - 1}}}}
\end{gather} 
where $Y(s)$ and $E_0(s)$ are the Laplace transforms of $y$ and $e_0$, respectively.
The open-loop transfer function of the IFO-ADRC system is
\begin{gather}
{G_{ifo}}(s) = \frac{{{Z_1}(s)}}{{E_0(s)}} = \frac{{{Z_1}(s)}}{{Y(s)}}\frac{{Y(s)}}{{E_0(s)}} \nonumber\\
\approx  {\bar{G}_{ifo}}(s)= k_p\frac{{\frac{1}{{{\omega _c}^{n - 1}}}}}{{{s^\chi }{{(\frac{1}{{{\omega _c}}}{s^\gamma } + 1)}^{n - 1}}}}
\label{key}
\end{gather}
where $Z_1(s)$ is the Laplace transform of $z_1$.
Letting $k_p = w^{\chi}_gw_c^{n-1}$ makes the approximation  $\bar{G}_{ifo}(s)$ as
 \begin{equation}
\bar{G}_{ifo}(s)=\frac{w^{\chi}_g}{s^\chi (Ts^\gamma+1)^{n-1}}
\end{equation}
 which is   a WBITF in (\ref{ifo-wbitf})
with  $\kappa = n-1$ and $T  = \frac{1}{\omega_c}$.
On the one hand, due to the existence of $n-1$ low-pass filter $1/(Ts^{\gamma} + 1)$, $\bar{G}_{ifo}(s)$  behaves like a BITF in the low-frequency band. On the other hand, for a larger $\omega_c$, the  approximate open-loop transfer function $\bar{G}_{ifo}(s)$ behaves more like
a BITF.
  
  The design principle of the IFO-ADRC is summarized as follows. First, the IFO-ESO is proposed and used to construct the controller (\ref{eq_8_}) to approximately convert the original uncertain plant into a cascade integer-order integrator in (\ref{eq_14}). Then, a tracking controller (\ref{eq_12}) is proposed to approximately shape the open-loop transfer function as a WBITF.

On the contrary, the high-order term $y^{(m)}$ is  included in the total disturbance for FO-ADRC in \cite{al2020robustness} and  further estimated by a higher-order ESO. As shown in Section \ref{sec:PA}, such a design of ESO has  worse estimation performance.
Moreover, less number of observer states and auxiliary tracking controller parameters can be used to achieve stable   IFO-ADRC than  FO-ADRC and IO-ADRC systems as demonstrated in Section \ref{sec:TDS}.

\subsection{Stablilty Analysis of IFO-ADRC}
In this section, the stability criteria for the ESO and the IFO-ADRC system are provided.
Let the observer error be
$ 
{e} = {x} - {z}.
$
From (\ref{eq_9}) and (\ref{eq_10}), the  equation of the extended state observer error can be written as
\begin{equation}
_0^CD_t^\textbf{q}e = Ae-Le_1 + Eh_{ifo}.
\label{eq_30_}
\end{equation}
The characteristic polynomial of the system (\ref{eq_30_}) can be obtained \cite{deng2007stability}:
\begin{equation}
\lambda (s) = {s^{(n-1)\gamma  + \chi +\nu }} + \sum\limits_{i = 1}^n {{\beta _i}} {s^{(n - 1)\gamma + \nu}} + {\beta _{n + 1}}
\label{eq_32_}
\end{equation}
Then,
Theorem \ref{th_IFO-ESO} will present the  bounded-input bounded-output (BIBO) stability of the error system (\ref{eq_30_}), when $h_{ifo}$ is bounded.  Theorem \ref{th_closed-loop} gives the BIBO condition  for 
the closed-loop system  using  characteristic equation method. 
A special case  when $m=n=2$ is elaborated in Theorem \ref{prop:stable}, showing that when the observer gain  is  sufficiently large, the closed-loop system is BIBO.
The proofs of Theorem \ref{th_IFO-ESO}  and  \ref{prop:stable}  are given in the Appendix.
\bthm Consider the error dynamics of IFO-ESO (\ref{eq_30_}). Let   $\omega_0 > 1$ and ${\beta _i} = C_{n + 1}^i{\omega _0}^i$ for $i = 1,2, \cdots ,n+1$. If $h_{ifo}$ is bounded,  then the IFO-ESO is  BIBO stable, regarding  $h_{ifo}$ as the input and $e_1$ as the output.
\label{th_IFO-ESO}
\ethm

\bthm   Consider the IFO-ADRC closed-loop system  composed of (\ref{eq_7}),  (\ref{eq_8_}) and (\ref{eq_12}). 
Let $ (p_1,q_1)$,  $ (p_2,q_2)$ and $ (p_3,q_3)$ be pairs of coprime positive integers satisfying  $\chi = \frac{p_1}{q_1}$, $\gamma = \frac{p_2}{q_2}$, and $\nu = \frac{p_3}{q_3}$.
Define a polynomial 
\begin{align}
&P(w) = ({w^{{p_1}{q_2}{q_3}}}({w^{(n - 1){q_1}{p_2}{q_3}}} + \sum\limits_{i = 1}^{n - 1} {{k_{{d_i}}}{w^{(i - 1){q_1}{p_2}{q_3}}}} ) + {k_p})\nonumber \\
&\times({w^{(n - 1){q_1}{p_2}{q_3} + {q_1}{q_2}{p_3} + {p_1}{q_2}{q_3}}} + \sum\limits_{i = 1}^n {{\beta _i}} {w^{(n - i){q_1}{p_2}{q_3} + {q_1}{q_2}{p_3}}} \nonumber \\ &+ {\beta _{n + 1}})
+ ({k_p} + \sum\limits_{i = 1}^{n - 1} {{k_{{d_i}}}({w^{{p_1}{q_2}{q_3} + (i - 1){q_1}{p_2}{q_3}}}} \nonumber \\  &+ \sum\limits_{j = 1}^{i - 1} {{\beta _j}{w^{j{q_1}{p_2}{q_3}}}}  + {\beta _i}) + ({w^{{p_1}{q_2}{q_3} + (n - 1){q_1}{p_2}{q_3}}} \nonumber \\&+ \sum\limits_{j = 1}^{n - 1} {{\beta _j}{w^{j{q_1}{p_2}{q_3}}}}  + {\beta _n}))\times ({w^{{q_1}{q_2}{p_3}}}\sum\limits_{i = 0}^{m - 1} {{a_i}{w^{iq_1q_2q_3}}} )
\label{eq-19}
\end{align}
Let $\lambda = \frac{1}{{q_1}{q_2}{q_3}}$ and
 $w_i$ be  the $i$th root of the equation (\ref{eq-19}) for $i=1,\cdots,2mp_1p_2p_3+q_1q_2p_3$.
 If $b=b_0$, ${{k_p},{k_d}_{_1}, \cdots , {k_d}_{_{m - 2}}}$ and $\beta_1, \cdots, \beta_{n+1}$ are selected such that all the roots satisfy $|\mbox{arg}({w_i})| > \frac{\lambda\pi }{{2}}$, then the IFO-ADRC closed-loop system is BIBO stable,
  regarding $d$ as the input, and $r-y$ as the output. 
 Moreover, the tracking error $r(t)-y(t)$ converges to a small neighborhood of the origin as $t\rightarrow \infty$.
\label{th_closed-loop}
\ethm

\bthm	\label{prop:stable}
Consider the IFO-ADRC closed-loop system  composed of (\ref{eq_7}),  (\ref{eq_8_}) and (\ref{eq_12}) with
$m=n=2$. Suppose the plant 
(\ref{eq_7}) is stable or marginally stable, i.e., $a_1 \ge 0$ and $a_0 \ge 0$.
Let ${\beta _i} = C_{n + 1}^i{\omega _o}^i$ for $i = 1,2,3$, $k_p>0$, $k_{d_1} > 8$, and $\gamma<\nu<1$. Then, there always exists a constant $\omega_0 > 0$, such that the closed-loop system is BIBO stable.
Moreover, the tracking error $r(t)-y(t)$ converges to a small neighborhood of the origin as $t\rightarrow \infty$.
	\ethm

\section{Performance Analysis of IFO-ESO in Frequency- domain \label{sec:PA}}
In this section, we will  compare the performance of the IFO-ESO proposed in Section \ref{sec:ADRC} with the FO-ESO proposed in   \cite{al2020robustness}, whose structure for second-order plant is illustrated in Fig. \ref{str_FO}.
Note that the role of the ESO in the framework of ADRC  is  to estimate the uncertain dynamics and external disturbances  to improve the robustness of the system. 
If the IFO-ESO  can perfectly estimate $\upfrown{f}_{ifo}$, 
the IFO-ADRC  can  convert the original system  into a cascaded integer-order integrator $1/{s^m}$ in  (\ref{eq_14}). 
Similarly,   the FO-ADRC   can  convert the original system  into   a fractional-order system $1/s^{m+\chi-1}$ (also looking from $u_0$ to $y$).
Therefore, we are motivated to use the model difference between $Y(s)/U_0(s)$ and the ideal model ($1/{s^m}$ with IFO-ESO; $1/s^{m+\chi-1}$ with FO-ESO)  to assess the performance of the two ESOs.
We adopt mean square error between $Y(s)/U_0(s)$ and the ideal model in the frequency-domain  to evaluate how difference  the two models  are. The mean square error  (MSE) of two linear models is defined as 
\begin{gather}
\text{IFO-ESO:}{\Delta _{ifo}(\omega)} = 1 - (j\omega)^{m}P_{ifo}(j\omega) \nonumber \\
\text{FO-ESO:}{\Delta _{fo}(\omega)} = 1 - (j\omega)^{m+\chi - 1} P_{fo}(j\omega)
\end{gather}
where $P_{ifo}(s)$  and $P_{fo}(s)$ are transfer function from $u_0$ to $y$ for IFO-ESO and FO-ESO, respectively. 
The MSE   was used in \cite{richardson1982parameter} for the model  identification where the problem is re-casted into an optimal problem of minimizing the model difference between the  identified and ideal model in terms of the MSE. Therefore, the MSE can be used to evaluate model difference in the frequency-domain.

\begin{figure}[!ht]
	\centering
	\includegraphics[scale=0.50]{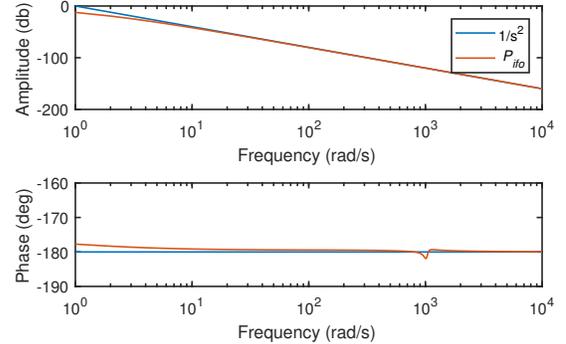}
	\caption{Bode plot of $P_{ifo}(j\omega)$  for  IFO-ESO.}
	\label{Pifo}
\end{figure}

\begin{figure}[!ht]
	\centering
	\includegraphics[scale=0.50]{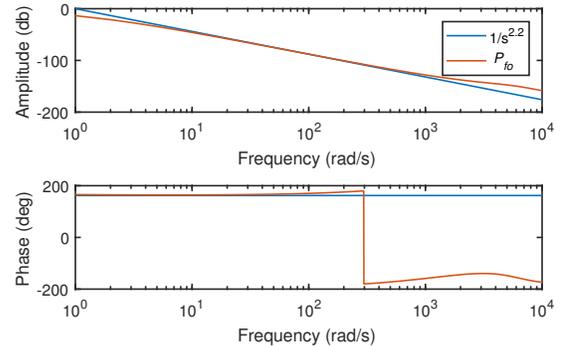}
	\caption{Bode plot of  $P_{fo}(j\omega) $ for  FO-ESO.  }
	\label{Pfo}
\end{figure}

\begin{figure}[!ht]
	\centering
	\includegraphics[scale=0.50]{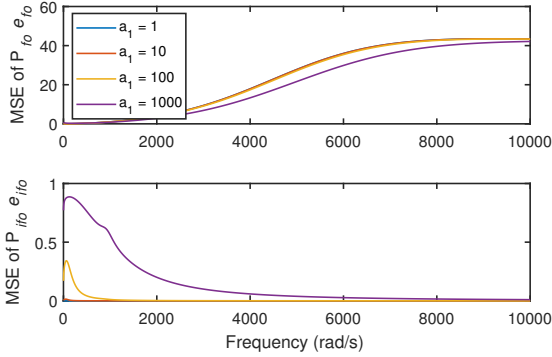}
	\caption{The MSE curves with different $a_1$ when $a_0 = 10$. 
	}
	\label{figure_5_2}
\end{figure}

\begin{figure}[!ht]
	\centering
	\includegraphics[scale=0.50]{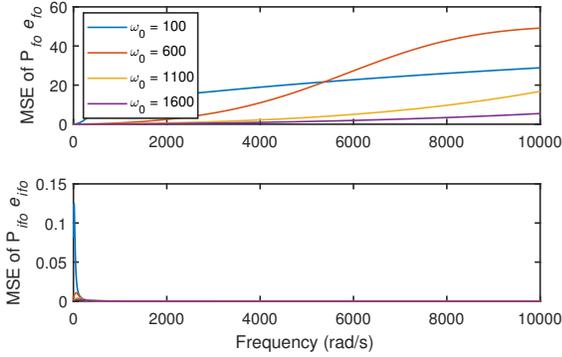}
	\caption{The MSE curves with different $\omega_0$ when $a_0 = 10$.
	}
	\label{figure_5_3}
\end{figure}

As in  \cite{al2020robustness}, for simplicity, we  consider the  second-order system 
\begin{equation} 
G(s) = \frac{{Y(s)}}{{U(s)}} = \frac{b}{{{s^2} + {a_1}s + {a_0}}},
\label{second-order system}
\end{equation}
 where the external disturbance and system uncertainty are not considered.
For the fair comparison, we choose the observer gains ${\beta _1} = 3{\omega _0}$, ${\beta _2} = 3{\omega _0}^2$, ${\beta _3} = {\omega _0}^3$  for IFO-ESO and ${\beta _1} = 4{\omega _0}$, ${\beta _2} = 6{\omega _0}^2$, ${\beta _3} = 4{\omega _0}^3$, ${\beta _3} = {\omega _0}^4$ for FO-ESO with $\omega_0 = 500$. The system parameters  are $b=5$, $a_0 = 10$, $a_1 = 10$. Let $\chi = 1.2$ for FO-ESO, $\nu = 0.8$, and $\gamma = 0.8$ for IFO-ESO.
The Bode plot of $P_{ifo}(j\omega)$  and $P_{fo}(j\omega)$    are illustrated in Fig.~\ref{Pifo} and Fig.~\ref{Pfo}, respectively.  It shows that the amplitude and the phase plot of  $P_{ifo}(j\omega)$ are close to  that of the ideal model within  a larger frequency band than that of $P_{fo}(j\omega)$. In particular, the approximation of $P_{fo}$ to $1/s^{2.2}$ gets worse in high-frequency band. 
It can be inferred that the IFO-ESO has better performance in terms  of disturbance estimation than the FO-ESO. 
Fig. \ref{figure_5_2} and Fig. \ref{figure_5_3}   show  the curves of the mean-square error $e_{fo}$ and $e_{ifo}$ with different model paramter $a_1$ and $\omega_0$, respectively,  when $a_0 = 10$. 
These two figures demonstrate that
 the mean-square error $e_{ifo}$  is less prone to the variation of system parameter $a_1$ and observer parameter $\omega_0$
  than   $e_{fo}$ is. In other words, the IFO-ESO can achieve better estimation performance than FO-ESO.
\section{Time-domain simulation and Comparison \label{sec:TDS}}
\begin{figure}[!ht]
	\centering
	\includegraphics[scale=0.55]{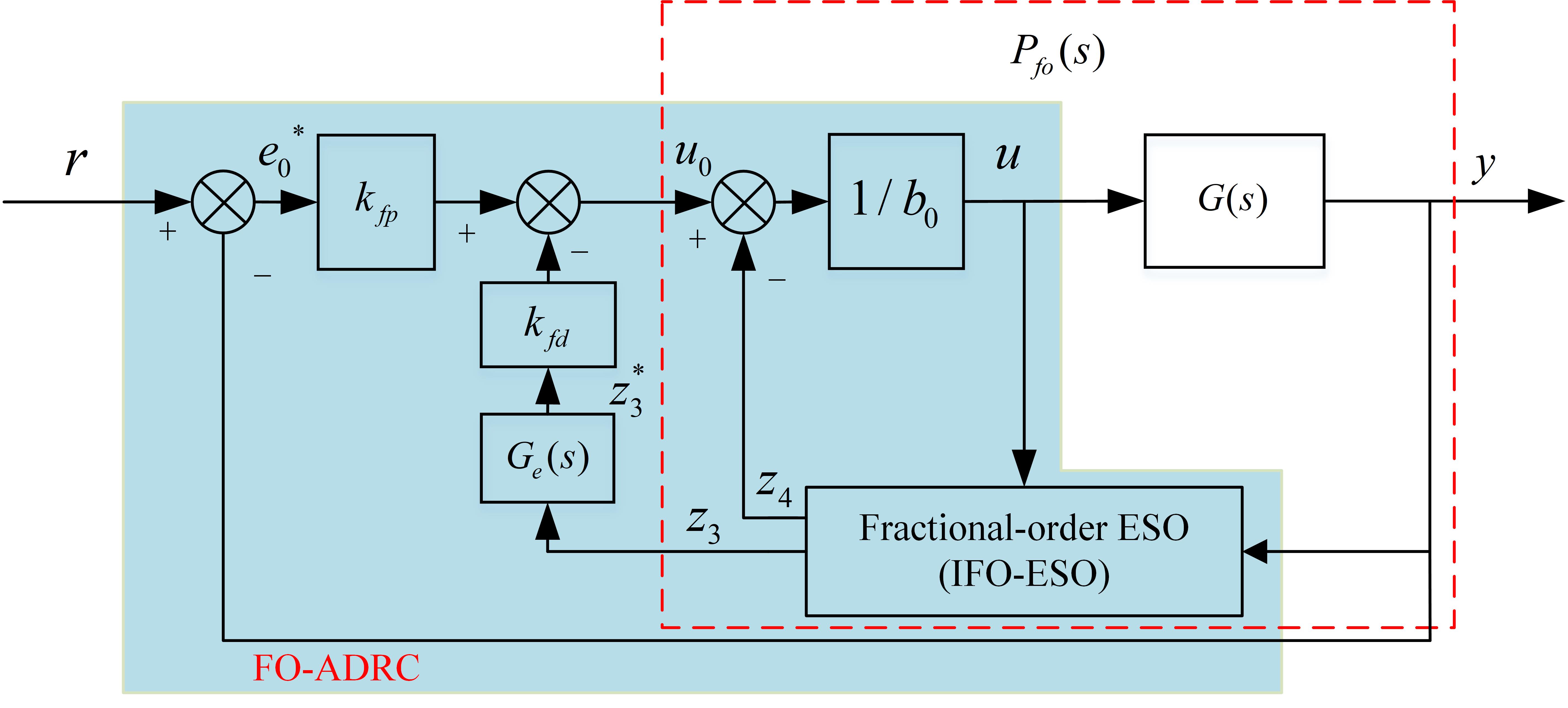}
	\caption{Structure of the FO-ADRC in \cite{al2020robustness} for a second-order plant.}
	\label{str_FO}
\end{figure}

\begin{figure}[!ht]
	\centering
	\includegraphics[scale=0.55]{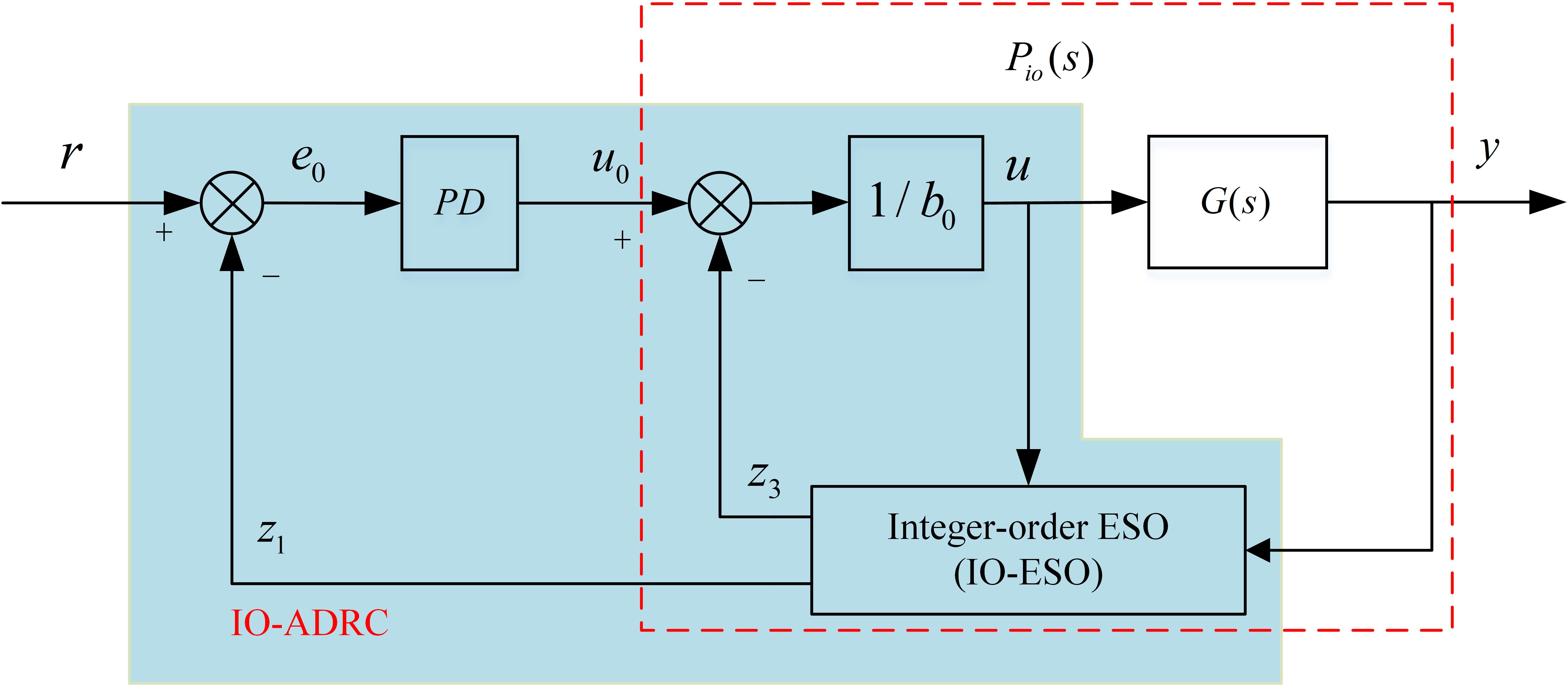}
	\caption{Structure of the IO-ADRC}
	\label{str_IO}
\end{figure}

In this section, we will show  the performance  of the IFO-ADRC in the time-domain using MATLAB/Simulink and compare it with IO-ADRC and FO-ADRC, respectively. The plant used for the simulation is a second-order system (\ref{second-order system}) where  $a_1 = 10$, $a_0 = 10$, $b_0 =b$.
  The structures of IFO-ADRC, FO-ADRC, and IO-ADRC with $\dot{r} = 0$ and $\ddot{r} =0$ are presented in Fig.~\ref{str_IFO}, \ref{str_FO}  and \ref{str_IO}, respectively. 
For IFO-ADRC, 
the observer gains $L=[\beta_1,\beta_2,\beta_3]\t=[3\omega_0,3\omega_0^2,\omega_0^3]\t$, $\omega_0 = 1200$, $\gamma = 0.8$, $\nu = 1.2$, and $\chi = 1.2$. The fractional-order  operators are discretized by the impulse response invariant method \cite{mis} where the discrete frequency for IFO-ESO is 8000 Hz and the discrete order of the fractional-order operators is 6. Note that the observer  parameters satisfy  conditions of  Theorem \ref{th_closed-loop}, the IFO-ADRC are BIBO stable.

The auxiliary controller $u_0$ in (\ref{eq_12}) for IFO-ADRC becomes 
\begin{gather}
{u_0} = {k_{ifp}}(r - {z_1}) - {k_{ifd}}{z_2}, \label{eq:u0_x}
\end{gather}
 with $k_p=k_{ifp} $, $k_{d_1}=k_{ifd} $. 
Accroding to (\ref{key}), the open-loop transfer function of the IFO-ADRC system can be approximately equivalent to weighted WBITF as follows
\begin{gather}
G_{ifo}(s) = \frac{{Z_1(s)}}{{E_0(s)}} \approx \frac{{k_{ifp}/k_{ifd}}}{{{s^{1.2}}(\frac{1}{{k_{ifd}}}{s^{0.8}} + 1)}}
\label{ifo-open}
\end{gather}

\subsection{Comparison with IO-ADRC}
\begin{figure}[t]
	\centering
	\includegraphics[scale=0.45]{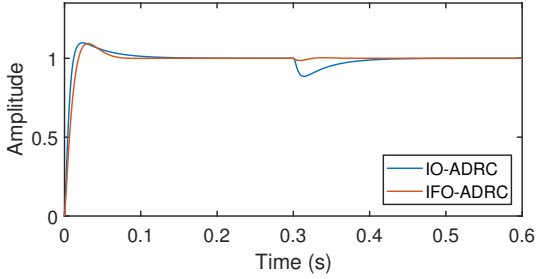}
	\caption{Step responses of two differemt control systems}
	\label{figure_4_2}
\end{figure}

For the fair comparison,
let $k_{ifp} $ and $k_{ifd}$ in (\ref{eq:u0_x}) be  $k_{ifp} = 1.2 \times 10^{6}$ and $k_{ifd} = 4000$ and the PD controller for  IO-ADRC be 
$
{C_{pd}}(s) = k_{ip}(1 + k_{id}s)
$ 
with $k_{ip}= 4466.16$ and $k_{id}=0.02562$ such that the open-loop transfer functions of IO-ADRC and IFO-ADRC have the  same gain crossover frequency   $\omega^*_c$ = 114 rad/s and  phase margin   ${\varphi _m}={71.3^ \circ}$.  
%
An external disturbance is imposed for some simulations at 0.3s and lasts until the end of the simulation.
The step responses of the IO-ADRC and IFO-ADRC systems are illustrated in Fig.~\ref{figure_4_2}, showing  that the IFO-ADRC  system has better dynamic response performance and disturbance rejection performance than the IO-ADRC system. 

\begin{table}[!t]
	\renewcommand{\arraystretch}{1.3}
	\caption{Comparison of the responses with three control systems (simulation)}
	\centering
	\includegraphics[scale=0.105]{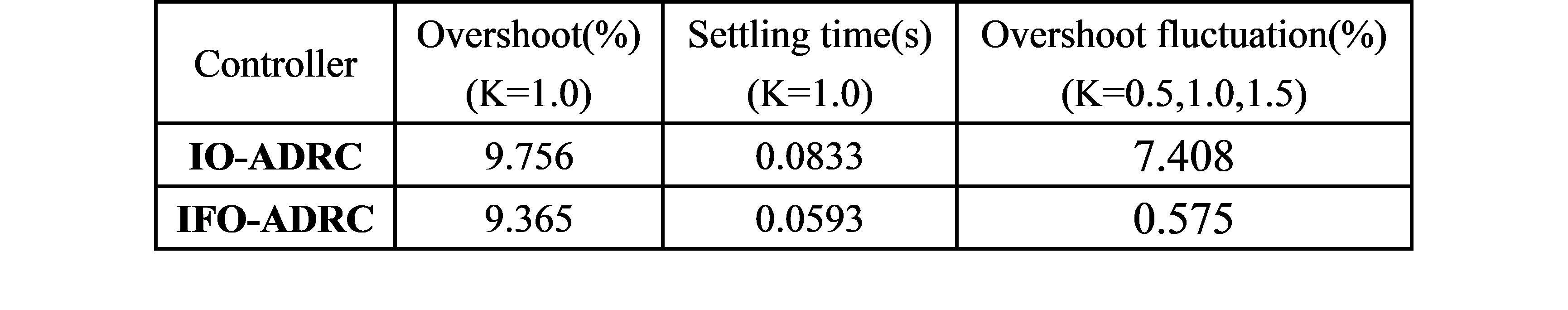}
	\label{table_1}
\end{table}

\begin{figure}[t]
	\centering
	\includegraphics[scale=0.45]{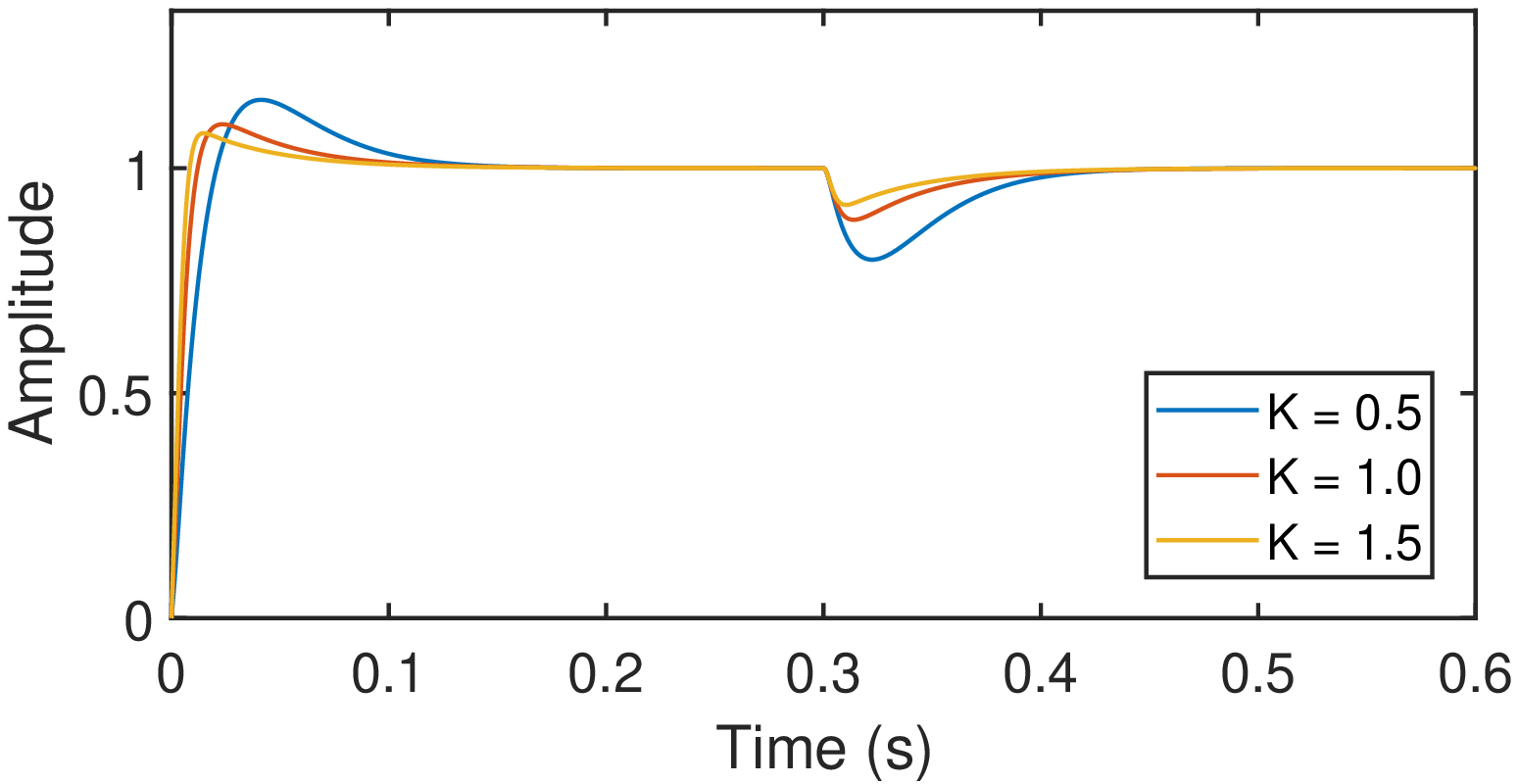}
	\caption{Step responses of the IO-ADRC control system with controller gain variations}
	\label{figure_5_1}
\end{figure}

\begin{figure}[t]
	\centering
	\includegraphics[scale=0.45]{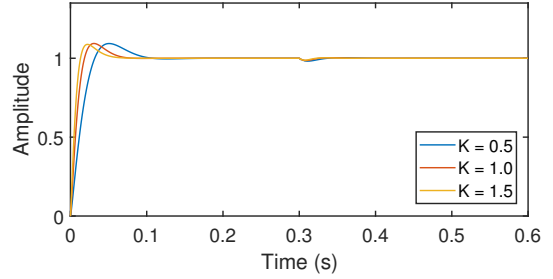}
	\caption{Step responses of the IFO-ADRC control system with controller gain variations}
	\label{figure_7_1}
\end{figure}

Now, let us consider the system performance  against 
controller parameters variation. Multiple the  controller parameter $K_{ifp}$ in the IFO-ADRC system and $K_{ip}$ in the IO-ADRC system by $K$ and consider cases with $K=0.5$, $K=1$ and $K=1.5$. 
Fig.~\ref{figure_5_1} and Fig.~\ref{figure_7_1} are the step responses of two closed-loop systems when different controller  parameters  are imposed. As shown in  Fig.~\ref{figure_7_1}, the IFO-ADRC  system are robust to controller gain variations.

When $K=0.5$, $K=1.0$, or $K=1.5$ are set respectively,  the maximum speed of the step response are denoted as $M_K$. The overshoot fluctuation is calculated as
$
\frac{{\max \{ {M_{0.5}},{M_{1.0}},{M_{1.5}}\}  - \min \{ {M_{0.5}},{M_{1.0}},{M_{1.5}}\} }}{\mbox{reference{\;\;} input}}.
$
The step responses of two  closed-loop systems with different $K$ are summarized in TABLE \ref{table_1}.  Note that the overshoots of IFO-ADRC system is smaller  than that of the IO-ADRC  system. The settling time of the IFO-ADRC system is shorter than the IO-ADRC system.r
\subsection{Comparison with FO-ADRC} 
For the fair comparison, 
let $k_{ifp} $ and $k_{ifd}$ in (\ref{eq:u0_x}) be $k_{ifp} = 9.6 \times 10^{4}$ and $k_{ifd} = 400$, and  the parameters of the FO-ADRC be $k_{fp} = 2.9328 \times 10^5$ and $k_{fd} = 1222$  (as illustrated in Fig. \ref{str_FO}). Let  the cut-off frequencies of the two low-pass filters be $\omega^o_c$ = 1222 rad/s. 
As a result, the  open-loop transfer function is approximated as a WBITF
\begin{gather}
{G_{fo}}(s)  \approx \frac{240}{{{s^{1.2}}(\frac{1}{{{1222}}}s + 1)}} \label{eq:gfo}
\end{gather} 
The two WBITFs (\ref{ifo-open}) and (\ref{eq:gfo})  have the  same gain and cut-off frequencies of the low-pass filters.

\begin{figure}[t]
	\centering
	\includegraphics[scale=0.45]{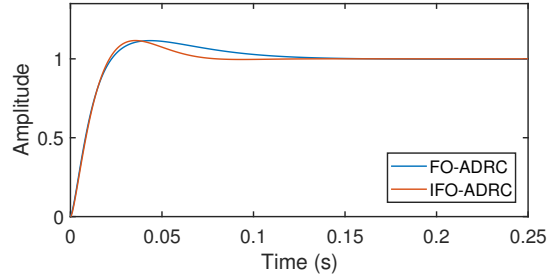}
	\caption{Step responses of the IFO-ADRC and FO-ADRC systems}
	\label{comp-fo}
\end{figure}

\begin{figure}[t]
	\centering
	\includegraphics[scale=0.45]{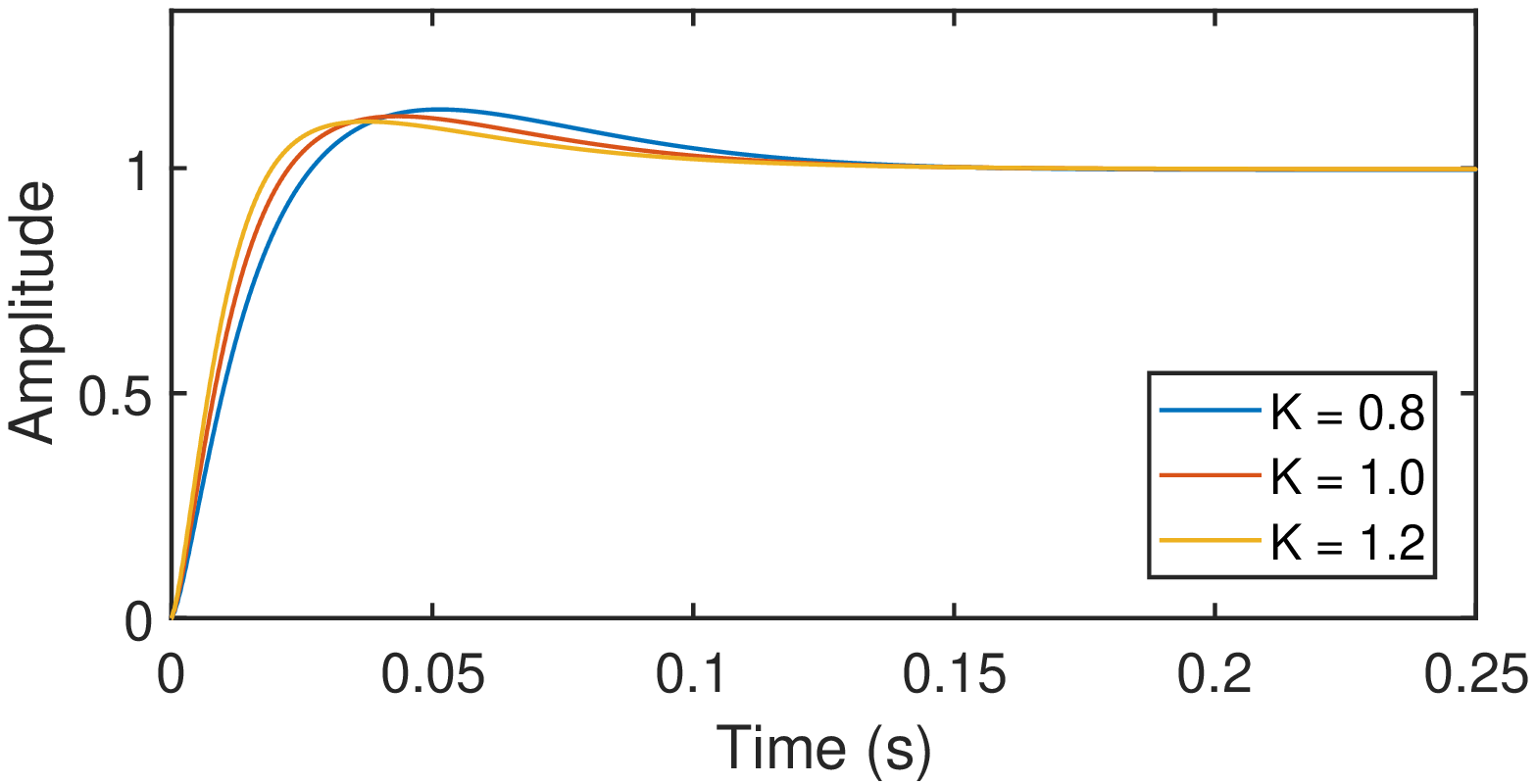}
	\caption{Step responses of the FO-ADRC systems with controller parameters varations}
	\label{fo-K}
\end{figure}

\begin{figure}[t]
	\centering
	\includegraphics[scale=0.45]{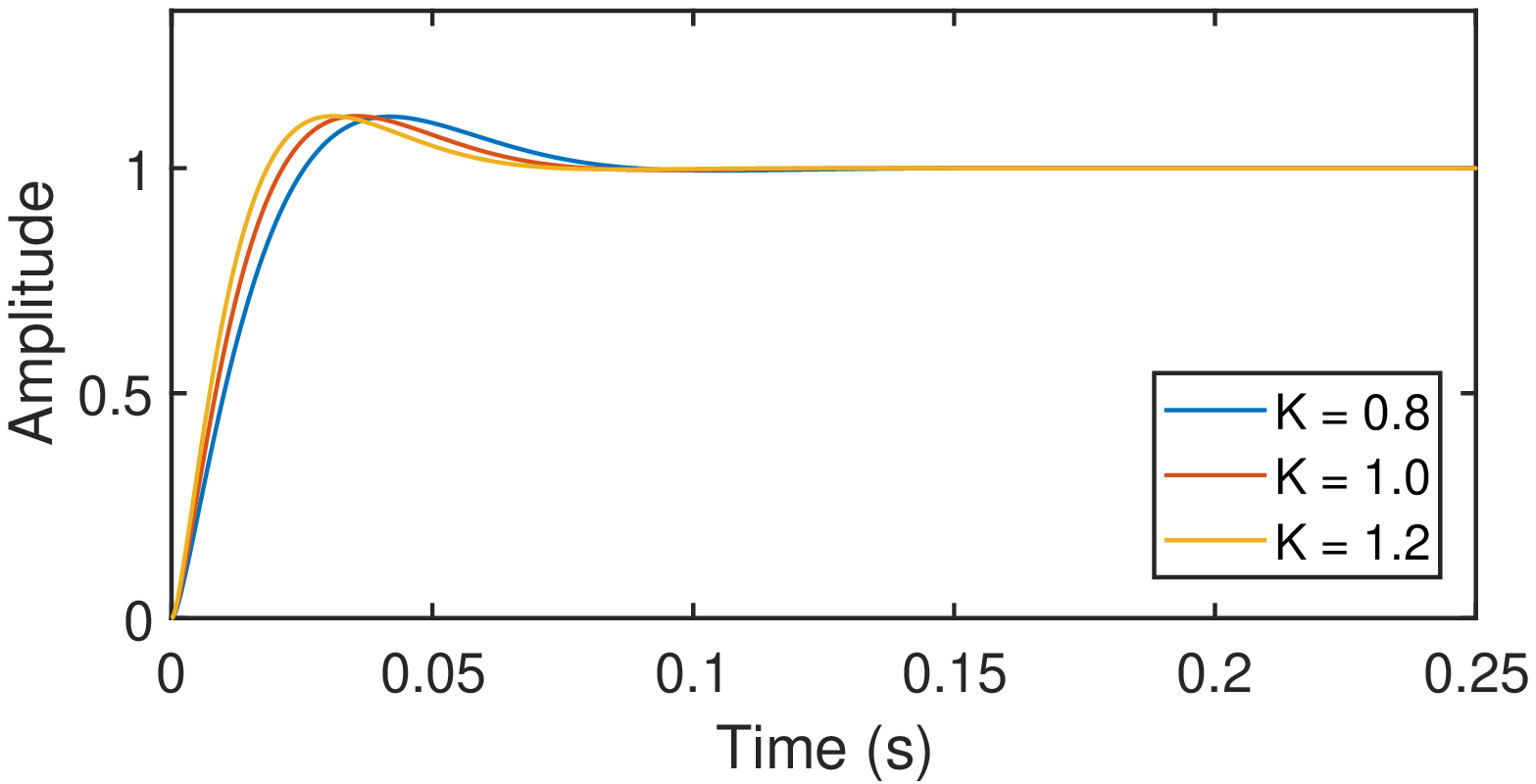}
	\caption{Step responses of the IFO-ADRC systems with controller parameters varations}
	\label{ifo-K}
\end{figure}

\begin{table}[!t]
	\renewcommand{\arraystretch}{1.3}
	\caption{Comparison of the responses with  the IFO-ADRC and FO-ADRC systems (simulation)}
	\centering
	\includegraphics[scale=0.105]{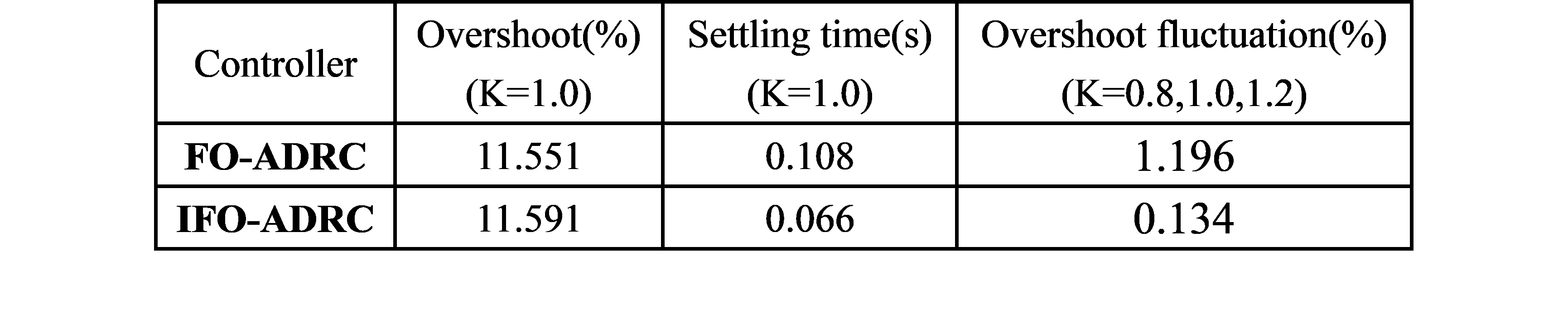}
	\label{COMP-FO}
\end{table}

 The step responses of the FO-ADRC and IFO-ADRC systems are illustrated in Fig.~\ref{comp-fo} showing that the step response of the IFO-ADRC system has shorter rise time, peak time, and settling time than that of the FO-ADRC system.  Multiple the  controller parameter $K_{ifp}$ in the IFO-ADRC system and $K_{fp}$ in the FO-ADRC system by $K$ and consider cases with $K=0.8$, $K=1$ and $K=1.2$. Fig.~\ref{fo-K} and Fig.~\ref{ifo-K} are the step responses of the IFO-ADRC and FO-ADRC systems when different controller  parameters are imposed.  It shows the IFO-ADRC system is more robust to controller parameters variations than FO-ADRC system. The step responses of the IFO-ADRC and FO-ADRC with different $K$ are summarized in TABLE \ref{COMP-FO}. It is clearly shown that the settling time of the IFO-ADRC system is shorter than the FO-ADRC system, although the overshoots of IFO-ADRC system is similar to that of the FO-ADRC  system.

\section{Experiments: PMSM speed servo control}
\begin{figure}[!ht]
	\centering
	\includegraphics[scale=0.45]{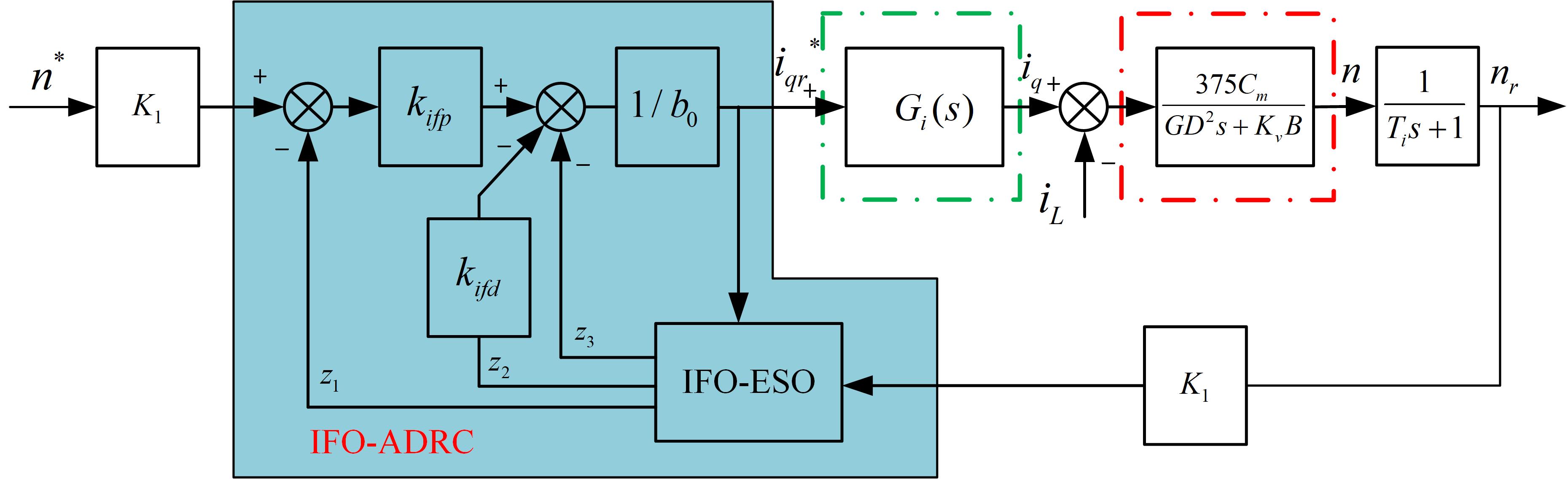}
	\caption{PMSM speed servo system using IFO-ADRC}
	\label{figure_21}
\end{figure}

In this section, the  performances of IFO-ADRC and IO-ADRC  are compared  on the PMSM speed servo control system experiments.
Fig. \ref{figure_21} is the block diagram of the speed loop of the PMSM speed servo system using IFO-ADRC. In Fig. \ref{figure_21}, $K_1$ is the speed conversion factor, $T_i$ is the speed feedback filter coefficient, and $n_r$ is the per unit of the actual speed. The block encircled by the green dash-dotted line in Fig.~\ref{figure_21} is the current loop $G_i(s)$ of the PMSM speed servo system. The PI controller in current loop is designed to ensure that $G_i(s) = 1$ in the operating frequency band of the speed loop. The plant of the PMSM speed servo system takes $i_q$ as the input and $n_r$ as the output.

\begin{figure}[!ht]
	\centering
	\includegraphics[scale=0.13]{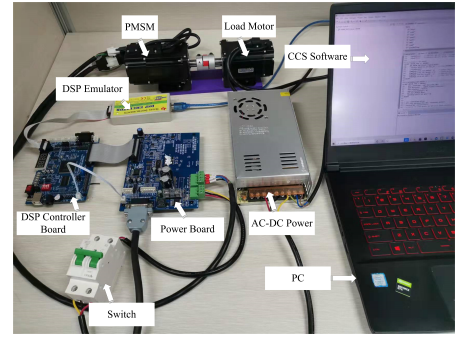}
	\caption{Experimental platform for control performance validation}
	\label{figure_9}
\end{figure}

The closed-loop controller is implemented on digital signal processor (DSP)  illustrated in Fig.~\ref{figure_9}. The PMSM is 60ST-M00630C and MOSFET is adopted as the gate driver.  
By system configuration, we  set $K_1 = 1/20000$, $T_i = 1/100$, $K_v = 375\pi C_m/30$, $C_m = 0.112$, $B = 5.747 \times 10^{-4}$ and $GD^2 = 1.539 \times 10^{-4}$. The speed sampling period was set as $1$ ms, and the current loop sampling period is set as $0.1$ ms. The motor speed waveform is collected by DSP Emulator and CCS software.
	When the specification of the PMSM is used,  the plant of the PMSM speed servo system becomes
	\begin{equation}
	G(s) = \frac{{1364.1}}{{{s^2} + 116.4s + 1642}}.
	\end{equation}
	The   same  controllers in Section \ref{sec:TDS} for IFO-ADRC  and IO-ADRC are used. Let the observer gain be $L=[\beta_1,\beta_2,\beta_3]\t=[3\omega_0,3\omega_0^2,\omega_0^3]\t$, ${\omega _0} = 700 $ rad/s, $b_0 = 1364.1$, $\chi = 1.2$, $\gamma=0.8$, $\nu$ = 1.2. The fractional-order  operators are discretized by the impulse response invariant method \cite{mis} where the discrete frequency for IFO-ESO is 1000 Hz and the discrete order of the fractional-order operators is 5.  Let $k_{ifp} = 9000$ and $k_{ifd} = 300$ for IFO-ADRC, 
	$k_{ip}= 266.255$ and $k_{id}=0.0854$ for IO-ADRC.
  For each experiment, a  constant load torque ($i_L \approx  0.5$ A) is imposed   at 0.75s and lasts until the end of the experiment to mimick the external  disturbance.


%
%

\begin{figure}[!ht]
	\centering
	\includegraphics[scale=0.45]{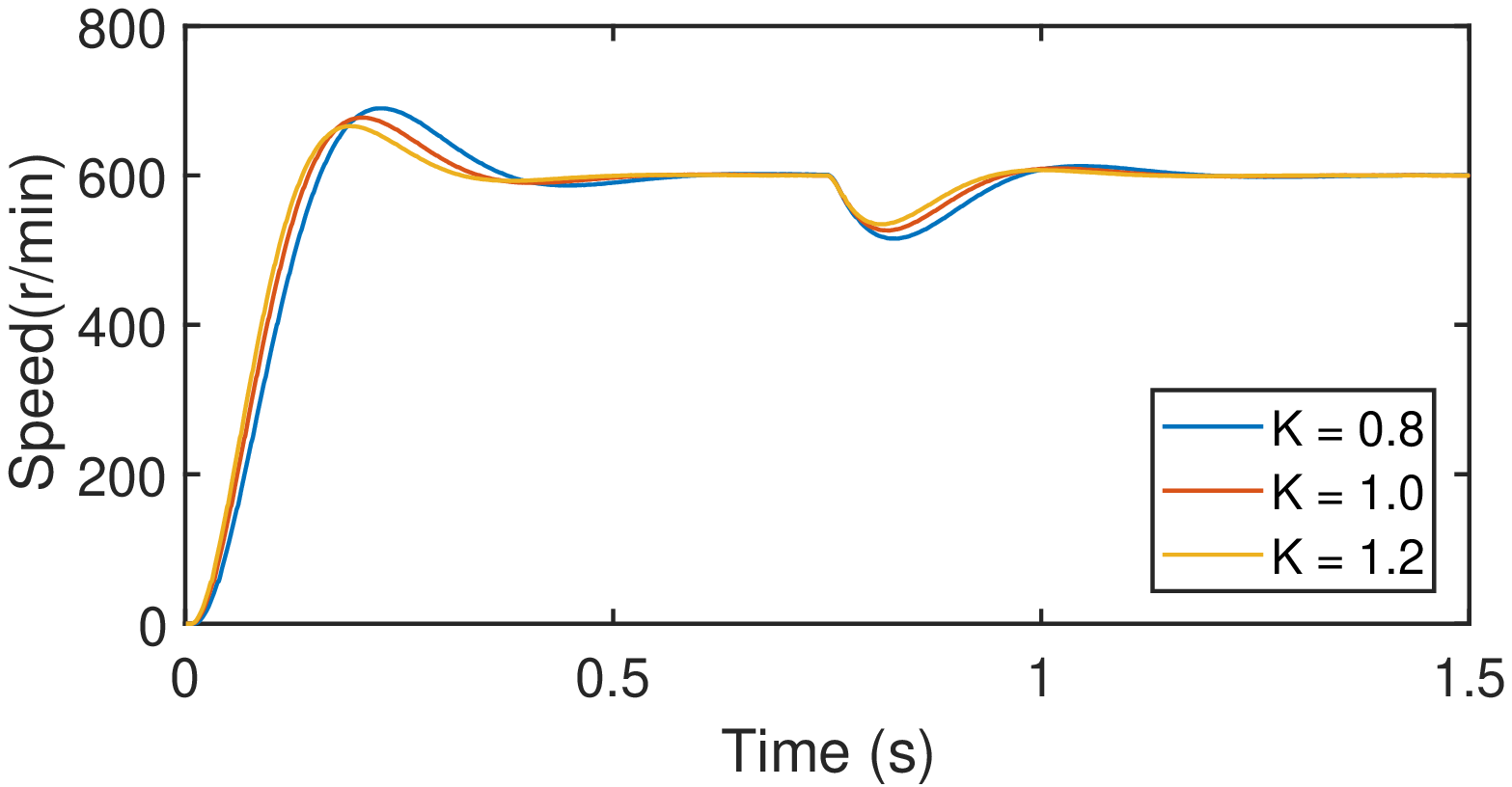}
	\caption{Step responses of the IO-ADRC with controller gain variations (experiment)}
	\label{figure_19}
\end{figure}

\begin{figure}[!ht]
	\centering
	\includegraphics[scale=0.45]{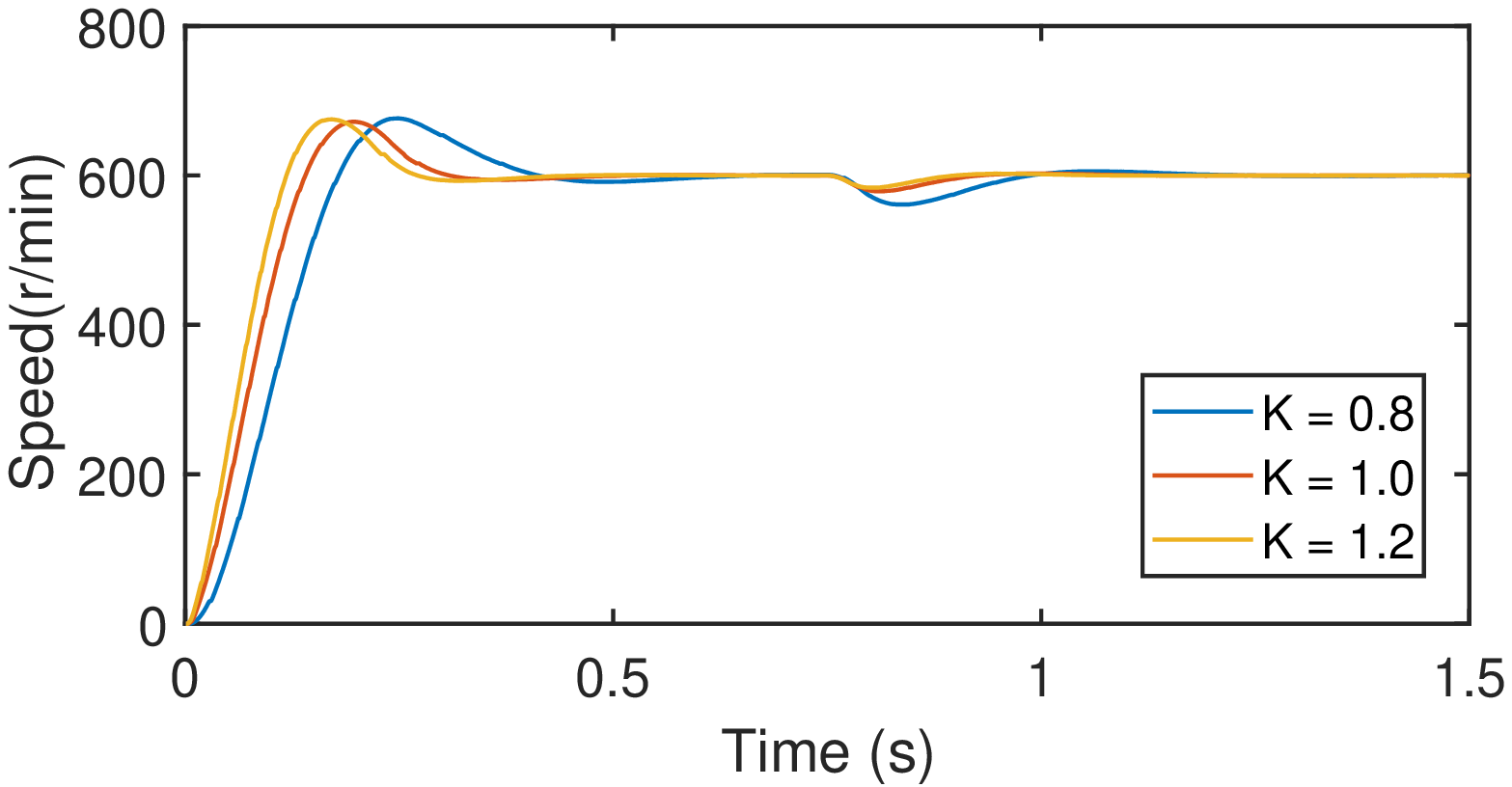}
	\caption{Step responses of the IFO-ADRC with controller gain variations (experiment)}
	\label{figure_16}
\end{figure}

\begin{figure}[!ht]
	\centering
	\includegraphics[scale=0.45]{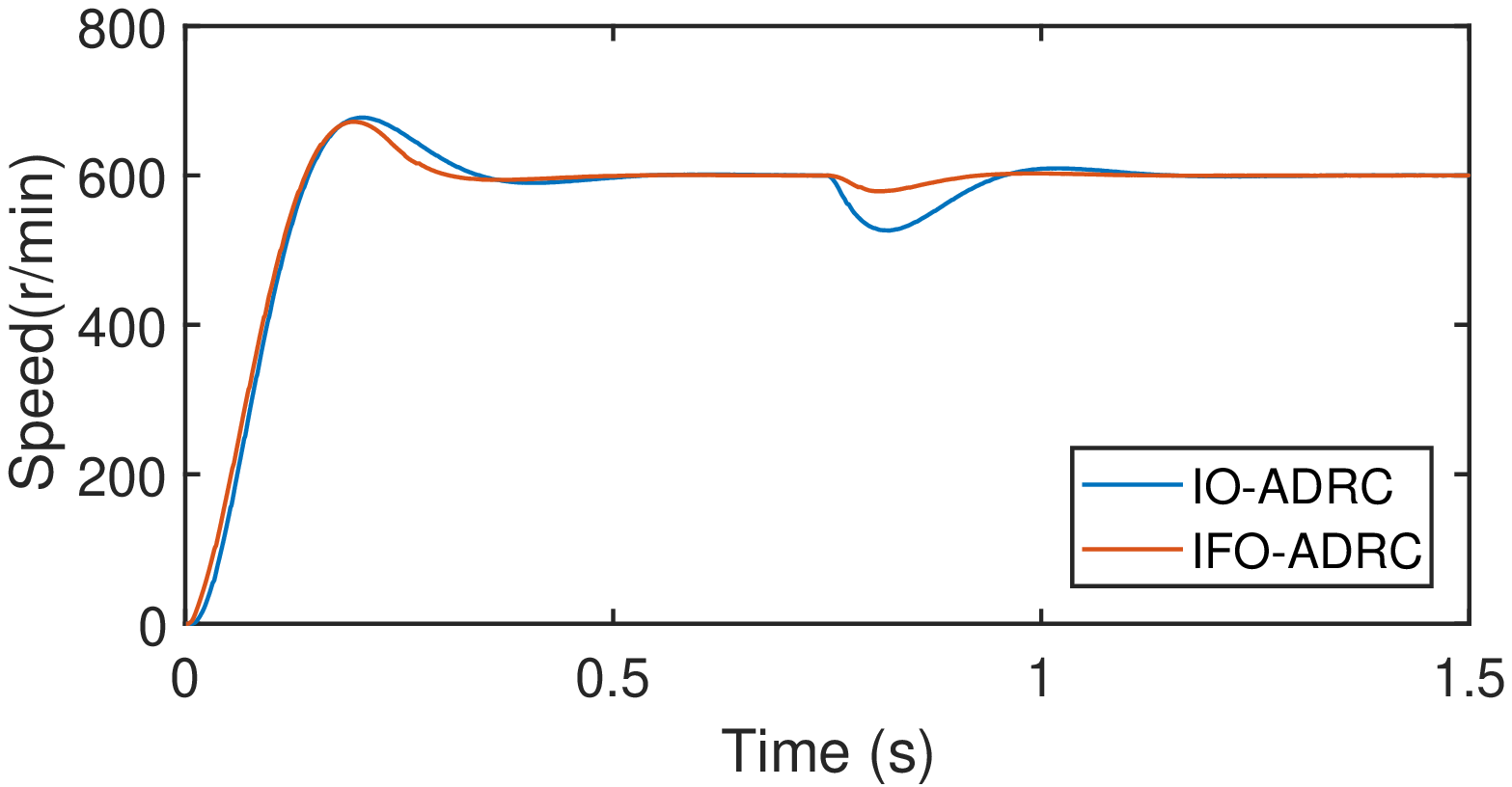}
	\caption{Step responses of two different control methods (experiment)}
	\label{figure_23}
\end{figure}

Fig.~\ref{figure_23} compares step responses of the IO-ADRC and IFO-ADRC systems, showing that the step response of the IFO-ADRC system has smaller overshoot, shorter settling time, and  smaller speed drop than the IO-ADRC system. 
Multiple the parameter $K_{ifp}$ in IFO-ADRC and PD controller parameter 
$K_{ip}$ in IO-ADRC by $K$ and consider $K=0.8$, $K=1$ and $K=1.2$. 
Fig.~\ref{figure_19} and Fig.~\ref{figure_16}  are experiment results of step responses with different $K$ 
for the IO-ADRC and IFO-ADRC systems, respectively. 
These two figures show that the IFO-ADRC system is robust to the open-loop gain variations. Table \ref{table_2} summarizes  results of the step responses for two different systems, showing that the IFO-ADRC system has better performance than the IO-ADRC system.

\begin{table}[!ht]
	\renewcommand{\arraystretch}{1.3}
	\caption{Comparison of the responses with three control systems (experiment) }
	\centering
	\includegraphics[scale=0.105]{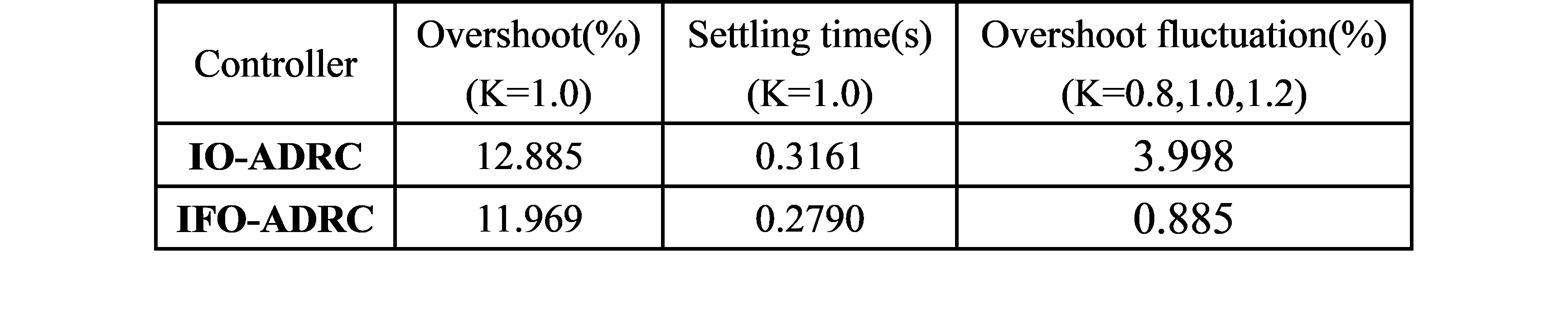}
	\label{table_2}
\end{table}

\section{Conclusion}
In this paper, an improved active disturbance rejection control scheme  is proposed  to approximately convert an integer-order system into a WBITF.  
The estimation performance of  IFO-ESO is better than
  FO-ESO, ensuring the closed-loop system is robust to  ESO parameters  and plant parameters variations.  
The frequency-domain analysis, time-domain simulation and PMSM speed servo control experiments verify that the proposed IFO-ADRC achieves better  performance than FO-ADRC and IO-ADRC. 

\appendix
\prooflater{Theorem \ref{th_IFO-ESO}}
Let $w = s^{\chi}$, and (\ref{eq_32_}) can be written:
\begin{equation}
\lambda (w) = {w^{(n-1)\sigma  + \delta + 1}} + \sum\limits_{i = 1}^n {{\beta _i}} {w^{(n - i)\sigma + \delta}} + {\beta _{n + 1}}
\label{eq_33_}
\end{equation}
where $\sigma  = \frac{\gamma }{{\chi}}$ and $\delta = \frac{\nu}{\chi}$.
Since $\gamma<\nu<\chi$, $n = [\frac{m}{\chi}]+1$, then $n - 1 > \frac{m}{\chi } - 1$, ie., $n\chi>m$ and $\sigma <\delta <1$. $\gamma = \frac{{m - \chi }}{n}$, thus $\chi > \gamma$, ie., $0 < \sigma  < 1$. 
According to Kharitonov-Based Method \cite{petravs2011fractional}, the three boundary polynomials (with (1) $\sigma$ = 0, $\delta$ = 0; (2) $\sigma$ = 0, $\delta$ = 1; (3) $\sigma$ = 1, $\delta$ = 1) are:
\begin{align}
^1\lambda (w) =& {w} + \sum\limits_{i = 1}^n {{\beta _i}} + {\beta _{n + 1}};\;\; ^2\lambda (w) = {w^2} + \sum\limits_{i = 1}^n {{\beta _i}}w + {\beta _{n + 1}};\nonumber \\
^3\lambda (w) =& {w^{n + 1}} + \sum\limits_{i = 1}^n {{\beta _i}} {w^{(n + 1 - i)}} + {\beta _{n + 1}}
\label{eq_34_}
\end{align}
Substituting ${\beta _i} = C_{n + 1}^i{\omega _0}^i$, $i = 1,2, \cdots ,n+1$ into ${}^3\lambda (w)$ gives ${}^4\lambda (w) = {(w + {\omega _0})^{n + 1}}$. Then,  the roots of the three boundary polynomials are:
\begin{align}
^1\lambda (w):{w_1} =&  - \sum\limits_{i = 1}^n {C_{n + 1}^i{\omega _0}^i}  - {\omega _0}^{n + 1}\nonumber \\
^2\lambda (w):{w_1} =& \frac{{ - \sum\limits_{i = 1}^n {C_{n + 1}^i{\omega _0}^i}  + \sqrt {{{(\sum\limits_{i = 1}^n {C_{n + 1}^i{\omega _0}^i} )}^2} - 4{\omega _0}^{n + 1}} }}{2}\nonumber \\
{w_2} =& \frac{{ - \sum\limits_{i = 1}^n {C_{n + 1}^i{\omega _0}^i}  - \sqrt {{{(\sum\limits_{i = 1}^n {C_{n + 1}^i{\omega _0}^i} )}^2} - 4{\omega _0}^{n + 1}} }}{2}\nonumber \\
{}^3\lambda (w):{w_i} =&  - {\omega _0},i = 1,2,3 \cdots ,n + 1
\label{eq_35_}
\end{align}
Since all the roots of the boundary polynomials are located in $|\mbox{arg}({w_i})| > \frac{\chi\pi }{{2}}$  when $\omega_0 > 1$ (all roots are located on the negative real axis), all the roots of (\ref{eq_33_}) are located in $|\mbox{arg}({w_i})| > \frac{\chi\pi }{{2 }}$ \cite{petravs2002robust}, i.e., $\lambda (s)$ is Hurwitz, system (\ref{eq_30_}) is BIBO stable, regarding $h_{ifo}$ as input and $e_1$ as output.\eproof

\begin{figure}[!ht]
		\centering
		\includegraphics[scale=0.60]{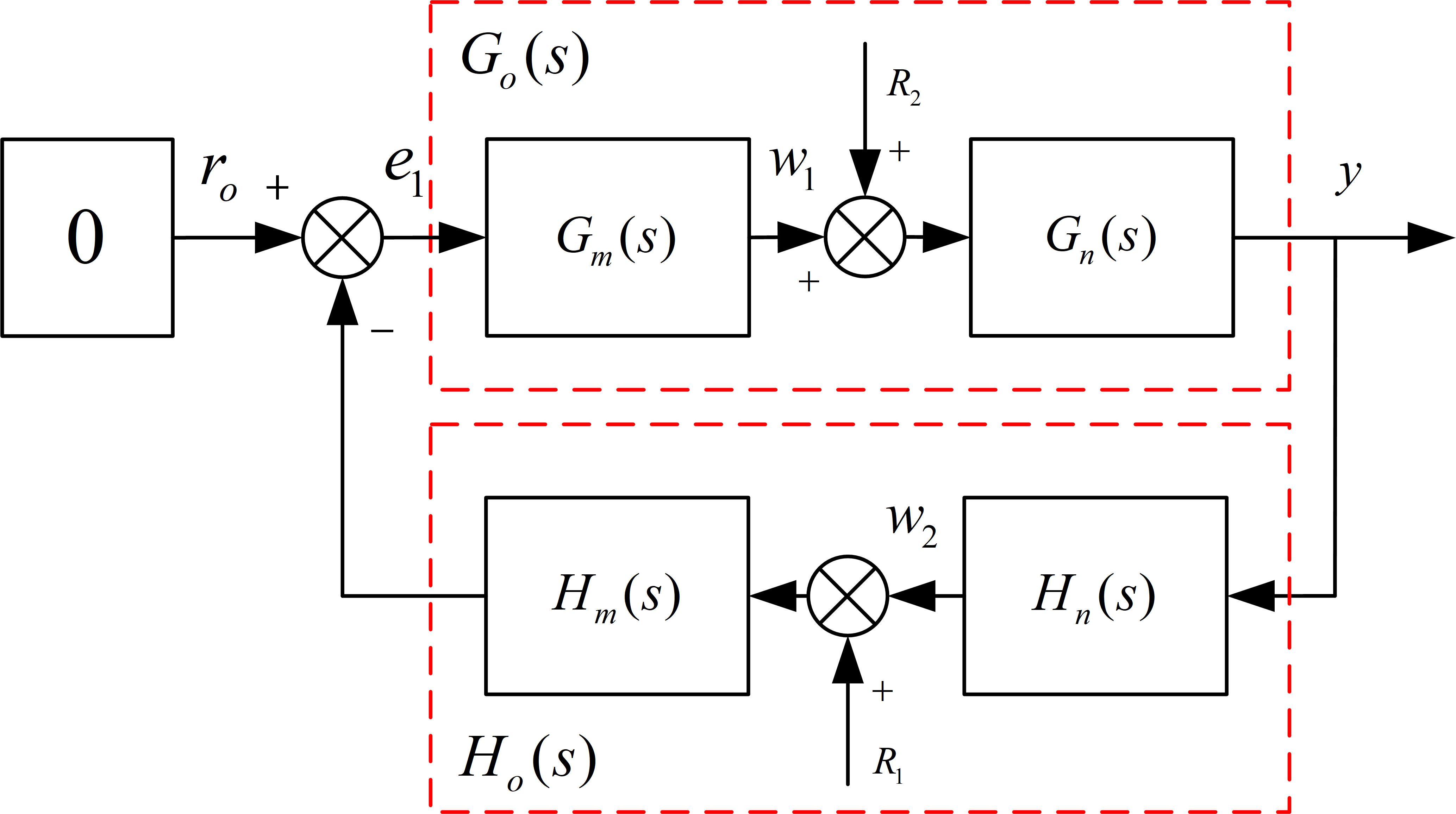}
		\caption{The block digram of the closed-loop system}
		\label{figure_24}
	\end{figure}

\prooflater{Theorem \ref{th_closed-loop}}
The system (\ref{eq_30_}) can be written
\begin{align} \label{eq:new_e}
{e_1}^{(\chi )} =&  - {\beta _1}{e_1} + {e_2} \nonumber \\
{e_2}^{(\gamma )} =&  - {\beta _2}{e_1} + {e_3}\nonumber\\
&\vdots \nonumber\\
e_n^{(\gamma)}=&   - {\beta _{n}}{e_1} + {e_{n + 1}}\nonumber\\
{e^{(\nu )}_{n + 1}} =&  - {\beta _{n+1}}{e_1} + {h_{ifo}}\nonumber\\
{h_{ifo}} =& -({a_0}{y}^{(\nu )} + {a_1}{y}^{(\nu + 1 )} + {a_2}{y}^{(\nu +2)} \nonumber \\
&+  \cdots  + {a_{m - 1}}{y}^{(\nu + m-1 )}) + R_1
\end{align} 
where
$
R_1 = d.$ It follows from (\ref{eq_14}), (\ref{eq_12}) and (\ref{eq_30_}) that
\begin{align}
{y^{(m)}} =& {{k_p}(r_1 - {x_1} + {e_1}) + {k_{{d_1}}}(r_2-{x_2} + {e_2}) +  \cdots}  \nonumber \\  +& {{k_{{d_{n - 1}}}}(r_n- {x_n} + {e_n}) + r_{n+1} - {x_{n + 1}} + {e_{n + 1}}}
\label{eq_41_}
\end{align}
It follows that
\begin{align}
{y^{(m)}} +& {k_{{d_{n - 1}}}}{x_n} +  \cdots  + {k_{{d_1}}}{x_2} + {k_p}{x_1} \nonumber \\ =& {k_p}(r_1 + {e_1}) + {k_{{d_1}}}{(r_2+e_2)} +  \cdots  + {k_{{d_{n - 1}}}}{(r_n + e_n)} \nonumber \\+& {r_{n+1}+e_{n + 1}}
\label{eq:new_62}
\end{align} 
Let $R_2 = {k_p}{r_1} + {k_{{d_1}}}{r_2} +  \cdots  + {k_{{d_{n-1}}}}{r_n} + {r_{n+1}}$, ${w_1}$ = ${k_p}{e_1} + {k_{{d_1}}}{e_2} +  \cdots  + {k_{{d_{n-1}}}}{e_n} + {e_{n+1}}$ and ${w_2} = -({a_0}{y}^{(\nu )} + {a_1}{y}^{(\nu + 1 )} + {a_2}{y}^{(\nu +2)} +  \cdots  + {a_{m - 1}}{y}^{(\nu + m-1 )})$. Combining (\ref{eq:new_e}) and (\ref{eq:new_62}) gives the block diagram of the closed-loop system in Fig. \ref{figure_24}.
From (\ref{eq:new_62}), it gives
\begin{align}
{G_m}(s) =  \frac{{{W_1}(s)}}{{E_1(s)}} = &{{k_p} + \sum\limits_{i = 1}^{n - 1} {{k_{{d_i}}}({s^{\chi  + (i - 1)\gamma }} + \sum\limits_{j = 1}^{i - 1} {{\beta _j}{s^{j\gamma }}}  + {\beta _i})}} \nonumber \\+& {({s^{\chi  + (n - 1)\gamma }} + \sum\limits_{j = 1}^{n - 1} {{\beta _j}{s^{j\gamma }}}  + {\beta _n})} \nonumber \\
{G_n}(s) = \frac{{Y(s)}}{{{W_1}(s)}} =& \frac{1}{{{s^\chi }({s^{(n - 1)\gamma }} + \sum\limits_{i = 1}^{n - 1} {{k_{{d_i}}}{s^{(i - 1)\gamma }}} )+k_p}} 
\end{align}
where $W_1(s)$, $Y(s)$ and $E_1(s)$ are the Laplace transforms of signals $w_1$, $y$ and $e_1$, respectively.
Accoding to (\ref{eq:new_e}), one has
\begin{align}
{H_m}(s) = &\frac{{{-E_1}(s)}}{{{W_2}(s)}} = \frac{-1}{{{s^{(n-1)\gamma  + \nu + \chi }} + \sum\limits_{i = 1}^n {{\beta _i}} {s^{(n - i)\gamma + \nu }} + {\beta _{n + 1}}}}\nonumber \\
{H_n}(s) = &  \frac{{{W_2}(s)}}{{Y(s)}} = - {s^\nu }\sum\limits_{i = 0}^{m - 1} {{a_i}{s^i}}    
\end{align}
where $W_2(s)$ is the Laplace transform of signals $w_2$.    
    
Now, let us show $R_2$ is bounded. Let ${d_0} = \sum\limits_{i = 0}^{m} {C_{m+1}^i{v^{m +1 - i}}{r^{(i)}}}$ where $v>0$ and $d_1 = {k_p}{r_1} + {k_{{d_1}}}{r_2} +  \cdots  + {k_{{d_{n-1}}}}{r_n}$. 
Define the transfer function of the system as
\begin{gather}
{P_1}(s) = \frac{{{D_1}(s)}}{{{D_0}(s)}} = \frac{{\sum\limits_{i = 1}^{n - 1} {{k_{{d_i}}}{s^{\chi  + (i - 1)\gamma }}}  + {k_p}}}{{{{({s} + v)}^{m + 1}}}}
\end{gather}
where $D_0(s)$ and $D_1(s)$ are the Laplace transforms of $d_0$ and $d_1$, respectively.
The characteristic polynominal of the system $P_1(s)$ is Hurwitz, ie., the system $P_1(s)$ is BIBO stable. As a result, $ \dot r,\ddot r, \cdots {r^{(m-1)}},{r^{(m)}}$ are  bounded, i.e., $d_0$ is bounded, then $d_1$ is bounded. Note that $R_2$ = $d_1+r^{(m)}$. Since $r^{(m)}$ is bounded,  $R_2$ is bounded.

Because $R_1$ and $R_2$ is bounded, we can treat $R_1$ and $R_2$ as the disturbance and calculate the transfer function of the closed-loop system. Further calculation gives    
\begin{align}
{G_o}{\rm{(s)  = }}\frac{{{Y}(s)}}{{E_1(s)}}{\rm{ = }}{{\rm{G}}_m}(s){G_n}(s)
\label{eq-64}
\end{align}
\begin{align}
{H_o}(s) = -\frac{{{E_1}(s)}}{{{Y}(s)}}{\rm{ = }}{H_m}(s){H_n}(s)
\label{eq-65}
\end{align}
Combining (\ref{eq-64}) and (\ref{eq-65}), the transfer function of the closed-loop system $P_o(s)$ can be given 
\begin{gather}
{P_o}(s) = \frac{{{Y}(s)}}{{{R_o}(s)}} = \frac{{{G_o}(s)}}{{1 + {G_o}(s){H_o}(s)}}
\label{closed-loop}
\end{gather}
where $R_o(s)$ is the Laplace transfoms of $r_o$ (see the Fig. \ref{figure_24}).
Finding positive integers $p_1$, $p_2$, $p_3$ $q_1$, $q_2$, and $q_3$  such that    $\chi = \frac{p_1}{q_1}$, $\gamma = \frac{p_2}{q_2}$, $\nu = \frac{p_3}{q_3}$.
the characteristic polynomial of the closed-loop system becomes (\ref{eq-19}).
 Since $R_1$ and $R_2$ is bounded, if the condition of Theorem \ref{th_closed-loop} is satisfied, the closed-loop system is BIBO stable \cite{petravs2011fractional}. 
 Moreover, the tracking error $q_1$ converges to a small neighborhood of the origin as $t\rightarrow \infty$.
\eproof

\prooflater{Theorem \ref{prop:stable}}
When $m=n=2$, the characteristic polynomial of the IFO-ADRC closed-loop system is
\begin{gather*}
P(s) = ({s^2} + {k_{d_1}}{s^\chi } + {k_p})({s^{2\gamma  + \chi }} + {\beta _1}{s^2} + {\beta _2}{s^\gamma } + {\beta _3}) \nonumber \\ + ({k_p} + {k_d}({s^\chi } + {\beta _1}) + {s^2} + {\beta _1}{s^{\gamma }} + {\beta _2})({s^\gamma }({a_0} + {a_1}s))
\label{char-poly}
\end{gather*}
According to Kharitonov-Based Method, since $\chi = 2 - \gamma$, $0<\gamma<1$, $\gamma<\nu<1$,  when ${\beta _i} = C_{n + 1}^i{\omega _0}^i$ for $i = 1,2,3$, the three boundary polynomials (with (1) $\gamma  = 0$, $\nu = 0$; (2) $\gamma=0$, $\nu = 1$; (3)$\gamma = 1$, $\nu = 1$)  can be written
\begin{align}
{}^1P(s) =& {{A_0}{s^4} + {A_1}s^3 + {A_2}}s^2+{A_3}s+{A_4}\nonumber \\
{}^2P(s) =& {{B_0}{s^5} + {B_1}{s^4} + {B_2}{s^3} + {B_3}{s^2} + {B_4}s + {B_5}}\nonumber \\
{}^3P(s) =& {{C_0}{s^5} + {C_1}{s^4} + {C_2}{s^3} + {C_3}{s^2} + {C_4}s + {C_5}}
\end{align}
with a list of parameters $A_i$, $B_i$ and $C_i$ for $i\in\{1,2,3,4,5\}$.
where
\begin{align}
{A_0} =& 1 + {k_{{d_1}}} + 3{\omega _0} + 3{k_{{d_1}}}{\omega _0}; 
{A_1} = {a_1} + {a_1}{k_{{d_1}}}\nonumber \\
{A_2} =& {a_0} + {a_0}{k_{{d_1}}} + {k_p} + 3{k_p}{\omega _0} + 3{\omega _0}^2 + 3{k_{{d_1}}}{\omega _0}^2 + {\omega _0}^3 \nonumber \\ +& {k_{{d_1}}}{\omega _0}^3\nonumber \\
{A_3} =& {a_1}{k_p} + 3{a_1}{\omega _0} + 3{a_1}{k_{{d_1}}}{\omega _0} + 3{a_1}{\omega _0}^2\nonumber \\
{A_4} =& {a_0}{k_p} + 3{a_0}{\omega _0} + 3{a_0}{k_{{d_1}}}{\omega _0} + 3{a_0}{\omega _0}^2 + 3{k_p}{\omega _0}^2 + {k_p}{\omega _0}^3\nonumber\\
{B_0} =& 1 + {k_{{d_1}}};
{B_1} = {a_1} + {a_1}{k_{{d_1}}} + 3{\omega _0} + 3{k_{{d_1}}}{\omega _0}\nonumber \\
{B_2} =& {a_0} + {a_0}{k_{{d_1}}} + {k_p} + 3{\omega _0}^2 + 3{k_{{d_1}}}{\omega _0}^2\nonumber \\
{B_3} =& {a_1}{k_p} + 3{a_1}{\omega _0} + 3{a_1}{k_{{d_1}}}{\omega _0} + 3{k_p}{\omega _0} + 3{a_1}{\omega _0}^2 + {\omega _0}^3 \nonumber \\+& {k_{{d_1}}}{\omega _0}^3\nonumber \\
{B_4} =& {a_0}{k_p} + 3{a_0}{\omega _0} + 3{a_0}{k_{{d_1}}}{\omega _0} + 3{a_0}{\omega _0}^2 + 3{k_p}{\omega _0}^2\nonumber\\
{B_5} =& {k_p}{\omega _0}^3;
{C_0} = 1; 
{C_1} = {a_1} + {k_{{d_1}}} + 3{\omega _0}\nonumber \\
{C_2} =& {a_0} + {a_1}{k_{{d_1}}} + {k_p} + 3{a_1}{\omega _0} + 3{k_{{d_1}}}{\omega _0} + 3{\omega _0}^2 \nonumber \\
{C_3} =& {a_0}{k_{{d_1}}} + {a_1}{k_p} + 3{a_0}{\omega _0} + 3{a_1}{k_{{d_1}}}{\omega _0} + 3{k_p}{\omega _0} + 3{a_1}{\omega _0}^2 \nonumber \\+& 3{k_{{d_1}}}{\omega _0}^2 + {\omega _0}^3\nonumber \\
{C_4} =& {a_0}{k_p} + 3{a_0}{k_{{d_1}}}{\omega _0} + 3a_0{\omega _0}^2 + 3kp{\omega _0}^2 + {k_{{d_1}}}{\omega _0}^3\nonumber \\
{C_5} =& {k_p}{\omega _0}^3
\label{eq-69}
\end{align}

Firstly, we consider ${}^1P(s) = 0$. When $a_1 \ge 0$, $a_0 \ge 0$, $k_p>0$, and $\omega_0 > 0$, from (\ref{eq-69}), we have $A_0>0$, $A_1 \ge 0$, and $A_2 > 0$. 
When $k_{d_1} > 8$ and $\omega_0$ is sufficiently large, $A_1A_2>A_3A_4$ and $A_1A_2A_3>A_0A_3^2+A_1^2A_4$. Accoring to Routh-Hurwitz criterion, all the roots of ${}^1P(s)=0$ are located in left plane, i.e., when $a_1 \ge 0$, ${a_0} \ge 0$ and $k_{d_1} > 8$, there exists a sufficiently large  $\omega_0 > 0$, such that all the root of ${}^1P(s)=0$ are located in left plane.

\begin{table}[!ht]
	\renewcommand{\arraystretch}{1.3}
	\caption{Routh table of ${}^2P(s) = 0$}
	\centering
	\includegraphics[scale=0.50]{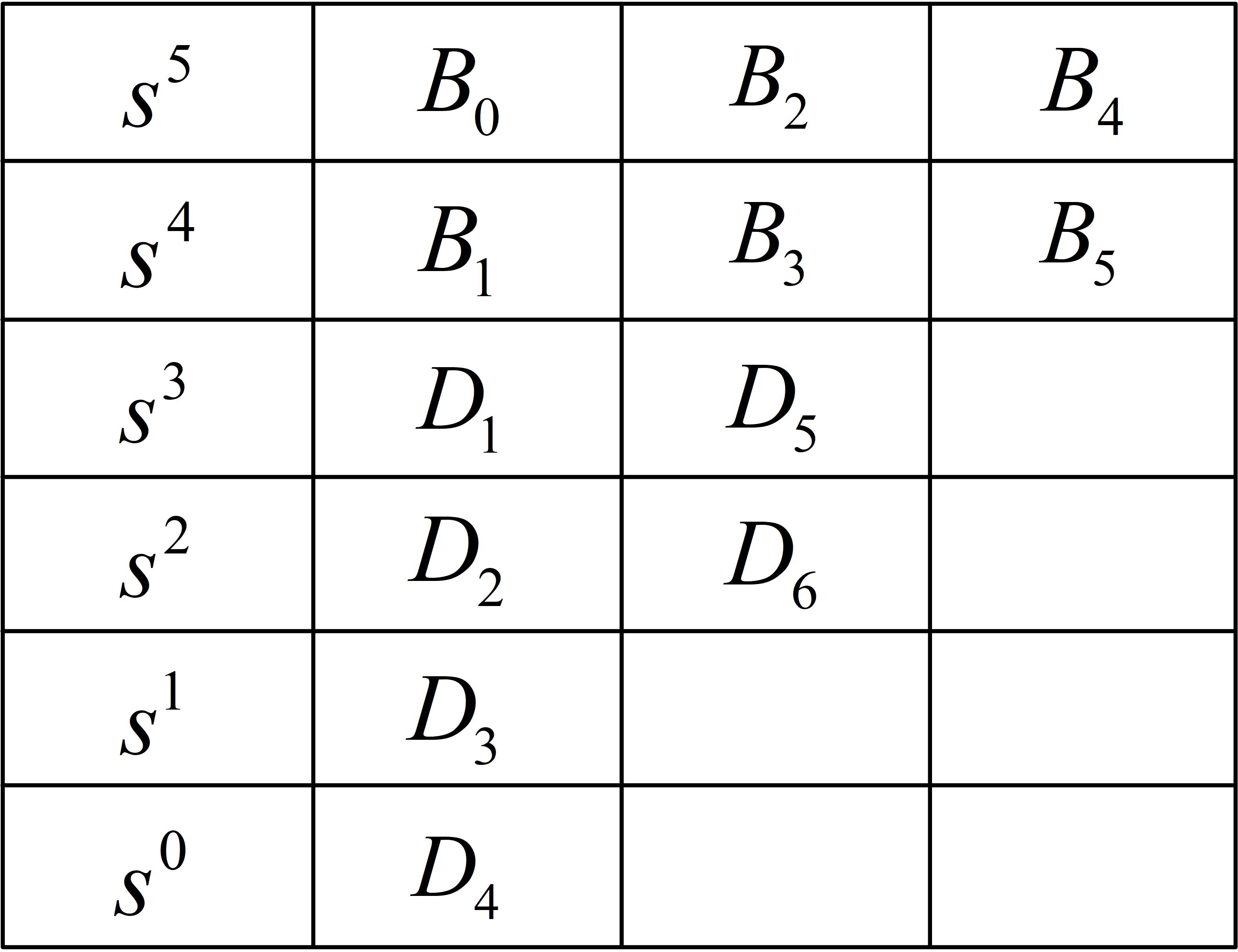}
	\label{table_4}
\end{table}
Secondly, we consider ${}^2P(s) = 0$. According to Routh-Hurwitz criterion, ${}^2P(s) = 0$ can be written as the Routh table form as TABLE \ref{table_4}. 
In TABLE \ref{table_4}, $C_1$, $C_2$, $C_3$, and $C_4$ are
\begin{equation}
{C_1} = \frac{{{N_5}}}{{{D_5}}},\;
{C_2} = \frac{{{N_6}}}{{{D_6}}},\;
{C_3} = \frac{{{N_7}}}{{{D_7}}},\;
{C_4} = k_p{\omega _o}^3
\end{equation}
where
\begin{align}
{N_5} =& {{a_0}{a_1} + 3{a_1}^2{\omega _o} + 9{a_1}{\omega _o}^2 + 8{\omega _o}^3},\;
{D_5} = {{a_1} + 3{\omega _o}}\nonumber \\
{N_6} =& 3{a_1}({a_0}^2 + {a_1}^2{k_p}){\omega _o} + (9{a_0}{a_1}^2 + 15{a_1}^2{k_p}){\omega _o}^2\nonumber \\ +& {a_1}(10{a_0} + 9{a_1}^2 + 18{k_p}){\omega _o}^3 + 30{a_1}^2{\omega _o}^4 + 33{a_1}{\omega _o}^5\nonumber \\ +& 8{\omega _o}^6 - 3{a_1}{a_0}{k_p} - 9{a_0}{k_p} - 3{a_0}{\omega _o}^4\nonumber \\
{D_6} =& {{a_0}{a_1} + 3{a_1}^2{\omega _o} + 9{a_1}{\omega _o}^2 + 8{\omega _o}^3\nonumber} \\
{N_7} =& 3({a_0}^3{a_1}{k_p} + {a_0}{a_1}^3{k_p}^2)+ (9{a_0}^2{a_1}^2{k_p} \nonumber \\ +& 15{a_0}{a_1}^2{k_p}^2){\omega _o} + (9{a_0}^3{a_1} + 9{a_0}^2{a_1}{k_p}\nonumber \\  +& 18{a_0}{a_1}^3{k_p} + 9{a_0}{a_1}{k_p}^2 + 9{a_1}^3{k_p}^2){\omega _o}^2 + (27{a_0}^2{a_1}^2\nonumber \\+& 96{a_0}{a_1}^2{k_p} - 24{a_0}{k_p}^2 + 42{a_1}^2{k_p}^2){\omega _o}^3 + (30{a_0}^2{a_1} \nonumber \\ +& 27{a_0}{a_1}^3 + 108{a_0}{a_1}{k_p} + 18{a_1}^3{k_p} + 48{a_1}{k_p}^2){\omega _o}^4 \nonumber \\ +& (90{a_0}{a_1}^2 + 54{a_1}^2{k_p}){\omega _o}^5 + (99{a_0}{a_1} + 48{a_1}{k_p}){\omega _o}^6\nonumber \\ +& 24{a_0}{\omega _o}^7 - 3{a_0}^2{a_1}{k_p}^2 - 9{a_0}^2{k_p}^2{\omega _o} - 30{a_0}^2{k_p}{\omega _o}^3 \nonumber \\-& 9{a_0}^2{\omega _o}^5 \nonumber \\
{D_7} =&
3{a_1}({a_0}^2 + {a_1}^2{k_p}) + (9{a_0}{a_1}^2 + 15{a_1}^2{k_p}){\omega _o}\nonumber \\ +& {a_1}(10{a_0} + 9{a_1}^2 + 18{k_p}){\omega _o}^2 + 30{a_1}^2{\omega _o}^3 + 33{a_1}{\omega _o}^4\nonumber \\ +& 8{\omega _o}^5 - 3{a_1}{a_0}{k_p} - 9{a_0}{k_p}\omega_o - 3{a_0}{\omega _o}^3
\label{eq-73}
\end{align}
When $a_0 \ge 0$, $a_1 \ge 0$, and $\omega_0$ is sufficiently large, we have $B_0 > 0$, $B_1> 0$, $C_1> 0$, $C_2> 0$, $C_3 > 0$, and $C_4> 0$.
According to Routh-Hurwitz criterion, if $B_0>0$, $B_1>0$, $C_1>0$, $C_2>0$, $C_3\ge0$, and $C_4>0$, then all the roots of ${}^2P(s) = 0$ are located in the left plane. Similar to the determination of the location of the root of ${}^2P(s) = 0$, when $a_0 \ge 0$, $a_1 \ge 0$, and $\omega_0$ is sufficiently large, all the roots of ${}^3P(s) = 0$ are located in the left plane.
According to Kharitonov-Based Method, when  the three boundary polynomials are stable, the closed-loop system is BIBO stable, ie., when $\gamma<\nu<1$, $m=n=2$, $a_1 \ge 0$, $a_0 \ge 0$,  $k_{d_1}> 8$ and  ${\beta _i} = C_{n + 1}^i{\omega _0}^i$ for $i = 1,2,3$,  there always exists a constant $\omega_0 > 0$, such that the closed-loop system is BIBO stable. 
\eproof

\bibliographystyle{IEEEtran}
\bibliography{./IEEEexample}
\vspace{-1.5cm}
\begin{IEEEbiography}[{\includegraphics[width=1in,height=1.25in,clip,keepaspectratio]{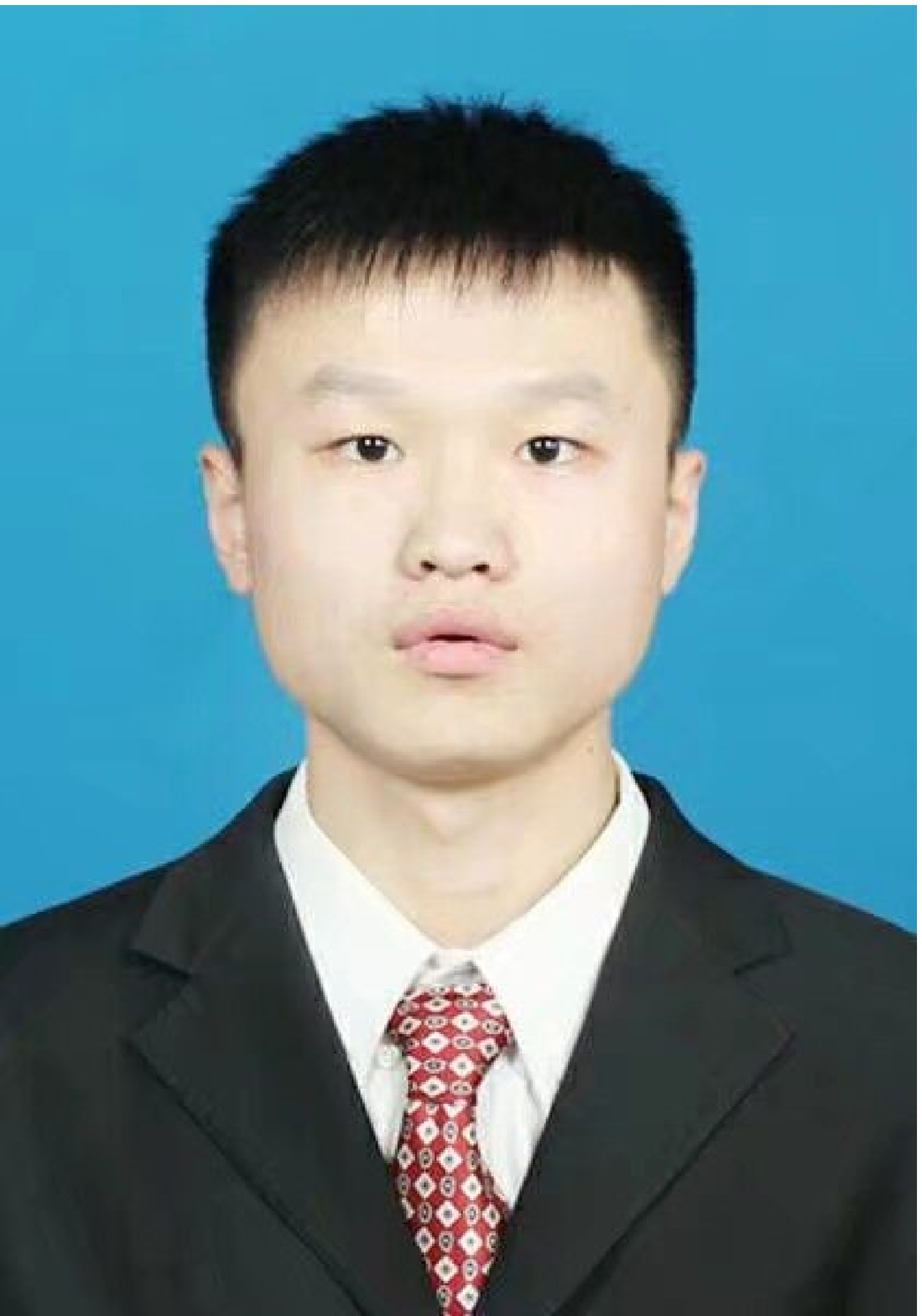}}]{Bolin Li}
received the M.E. degree in Mechanical and Electronic Engineering from Huazhong University of Science and Technology, Wuhan, China, in 2021. He is currently pursuing the Ph.D. degree
in School of Artificial Intelligence and Automation, the Huazhong University of Science and Technology (HUST), Wuhan, China. His research mainly involves active disturbance rejection control and fractional-order PID control.  
\end{IEEEbiography}

\vspace{-2cm}
\begin{IEEEbiography}[{\includegraphics[width=1in,height=1.25in,clip,keepaspectratio]{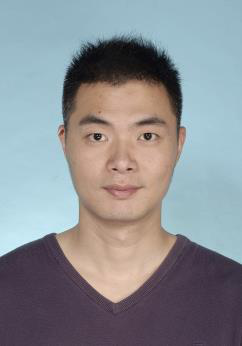}}]
	{Lijun Zhu}  
	 received
	the Ph.D. degree in Electrical Engineering from University of Newcastle,
	Callaghan, Australia, in 2013. He is now a Professor in  the School of Artificial Intelligence and Automation, Huazhong University of Science and Technology, Wuhan, China.
	His research interests include robotics,  
	nonlinear systems analysis and control.
\end{IEEEbiography}

%
\end{document}